\documentstyle[12pt]{article}
\def\ga{\mathrel{\mathpalette\fun >}}
\def\fun#1#2{\lower3.6pt\vbox{\baselineskip0pt\lineskip.9pt
\ialign{$\mathsurround=0pt#1\hfil##\hfil$\crcr#2\crcr\sim\crcr}}}

\newcommand{\be}{\begin{eqnarray}}
\newcommand{\ee}{\end{eqnarray}}
\newcommand{\ba}{\begin{array}}
\newcommand{\ea}{\end{array}}

\input{epsfig.sty}

\begin{document}

\Huge{\noindent{Istituto\\Nazionale\\Fisica\\Nucleare}}

\vspace{-3.9cm}

\Large{\rightline{Sezione SANIT\`{A}}}
\normalsize{}
\rightline{Istituto Superiore di Sanit\`{a}}
\rightline{Viale Regina Elena 299}
\rightline{I-00161 Roma, Italy}

\vspace{0.65cm}

\rightline{INFN-ISS 97/8}
\rightline{June 1997}

\vspace{1.5cm}

\begin{center}

\Large{EXCLUSIVE AND INCLUSIVE WEAK DECAYS\\ OF THE $B$-MESON\footnote{
{\bf To appear in Nuclear Physics B.}}}

\vspace{1cm}

\large{I.L. Grach$^{(1)}$, I.M. Narodetskii$^{(1)}$, S. Simula$^{(2)}$
and K.A. Ter-Martirosyan$^{(1)}$}

\vspace{0.5cm}

\normalsize{$^{(1)}$Institute for Theoretical and Experimental Physics,
117259 Moscow, Russia \\ $^{(2)}$Istituto Nazionale di Fisica Nucleare, 
Sezione Sanit\`{a}, \\ Viale Regina Elena 299, I-00161 Roma,
Italy}

\end{center}

\vspace{1cm}

\begin{abstract}

\noindent A relativistic quark model is applied to the description of
semileptonic and non-leptonic charmed decays of the $B$-meson. The
exclusive semileptonic modes $B \to D \ell \nu_{\ell}$ and $B \to D^*
\ell \nu_{\ell}$ are described through the universal Isgur-Wise form
factor, which is calculated in terms of a constituent quark model wave
function for the $B$-meson. Different approximations for the latter,
either based on a phenomenological ans\"{a}tz or derived from analyses
of the meson spectra, are adopted. In particular, two wave functions,
constructed via the Hamiltonian light-front formalism using a relativized
and a non-relativistic constituent quark model, are considered,
obtaining a link between standard spectroscopic quark models and the
$B$-meson decay physics. The inclusive semileptonic and non-leptonic
branching ratios are calculated within a convolution approach, inspired
by the partonic model and involving the same $B$-meson wave function
used for the evaluation of the exclusive semileptonic modes. Our results
for the major branching ratios are consistent with available
experimental data and the sum of all the calculated branching ratios
turns out to be close to unity. In particular, we found that: ~ i) a
remarkable fraction ($\sim 35 \%$) of semileptonic decay modes occur in
non-$D$, non-$D^*$ final states; ~ ii) non-perturbative effects enhance
the inclusive $b \to c \bar{u}d$ decay channels, with a sizable
contribution provided by {\em internal} decays into heavy mesons and
baryon-antibaryon pairs; the resulting reduction of the semileptonic
branching ratio brings the theoretical prediction in agreement with the
experimental value without increasing at the same time the charm
counting.

\end{abstract}

\newpage

\pagestyle{plain}

\section{Introduction.}

\indent The investigation of semileptonic ($SL$) and non-leptonic ($NL$)
decays of the $B$-meson can provide relevant information on the
fundamental parameters of the Standard Model of the electroweak
interaction and on the internal structure of hadrons. However, the
extraction of the Cabibbo-Kobayashi-Maskawa ($CKM$) \cite{CKM} mixing
parameters from the experiments requires a precise knowledge of the
transition form factors relevant in the matrix elements of the weak
hadronic current. In particular, the most accurate determination of the
$CKM$ matrix element $|V_{bc}|$ is presently based on the value of the
$B \to D^* \ell \nu_{\ell}$ transition form factors near the point of
zero recoil \cite{DAN94,BES94,SKW95,NEU96}, while the evaluation of the
lepton spectra and branching ratios requires the knowledge of the
transition form factors in their full kinematical range.

\indent As far as the theoretical point of view is concerned, the Heavy
Quark Effective Theory ($HQET$) \cite{HQET} is widely recognized as a
very powerful tool for investigating the decay modes of heavy flavours
and, recently \cite{OPE}, a model-independent framework has been
developed for the treatment of inclusive decays. The latter approach
relies on the formalism of $HQET$ and on the use of the operator product
expansion ($OPE$) in the physical region of time-like momenta. Thus, the
hypothesis of quark-hadron duality, in its global form for $SL$ decays
and in the local one for $NL$ processes, has to be invoked \cite{PQW76}.
The concept of quark-hadron duality, though it has not yet been derived
from first principles, is essential in the $QCD$ phenomenology and
corresponds to the assumption that the sum over many hadronic final
channels eliminates bound-state effects related to the specific
structure of each individual final hadron. The validity of the global
duality has been tested in inclusive hadronic $\tau$ decays \cite{GN96},
whereas the possibility of a failure of the local duality in inclusive
$NL$ processes has been recently raised in Ref. \cite{AMPR96}. It should
be reminded that the model-independent approach of Ref. \cite{OPE} has
other limitations, like, e.g., the need of a (at least partial) $OPE$
resummation in the end-point region of the predicted lepton spectrum.
Moreover, though the $HQET$ provides a systematic expansion for
organizing power corrections, it does not help in calculating the
relevant (non-perturbative) hadronic matrix elements. Therefore, the use
of the phenomenological constituent quark ($CQ$) model (see, e.g.,
\cite{ISGW,BSW,CW94}) for the description of the non-perturbative
aspects of the hadron structure could be still of interest. In this
respect, it is well known that the $CQ$ model is remarkably successful
in describing the non-perturbative physics of hadron mass spectra;
however, a successful model of hadrons must go beyond the spectroscopy
and should describe the internal structure of hadrons in order to
predict, e.g., transition form factors and decay rates.

\indent In this paper non-perturbative $QCD$ effects are mocked up by a
$CQ$ model wave function for the $B$-meson. The internal motion of the
$b$-quark inside the $B$-meson\footnote{The effects due to the
bound-state structure have been firstly treated in Ref. \cite{ACCMM} by
attributing a Fermi motion to the $b$-quark in the $B$-meson.} is
described by a distribution function $|\chi(x, p_{\perp})^2|^2$, which
represents the probability to find the $b$-quark carrying a light-front
($LF$) fraction $x$ of the $B$-meson momentum and a transverse relative
momentum squared $p_{\perp}^2 \equiv |\vec{p}_{\perp}|^2$. The branching
ratios of all the major decay modes of the $B$-meson are calculated in
terms of the distribution $|\chi(x, p^2_{\perp})|^2$. Thus, a relevant
feature of our approach is that both exclusive and inclusive $SL$ as
well as $NL$ decays modes are coherently treated in terms of the same
$b$-quark distribution in the $B$-meson. The sensitivity of our
predictions to various choices of $\chi(x, p^2_{\perp})$, based on the
phenomenological ans\"{a}tz of Ref. \cite{MT96} or derived from analyses
of the meson spectra, is investigated. As to the latter case, two wave
functions, constructed via the Hamiltonian $LF$ formalism using a
relativized \cite{GI85} and a non-relativistic \cite{NCS92} $CQ$ model,
have been considered, obtaining a link between standard spectroscopic
quark models and the $B$-meson decay physics.

\indent As it is well known, $B$-meson decays occur mainly through the
$CKM$ favored $b \to c$ transition. The dominant diagrams are the
so-called spectator diagrams depicted in Figs. 1 and 2. In $SL$ decays
(Fig. 1) the virtual $W^-$ boson materializes into a lepton pair, and
the $c$-quark and the spectator light antiquark hadronize independently
of the leptonic current. In case of the exclusive $SL$ transitions $B
\to D \ell \nu_{\ell}$ and $B \to D^* \ell \nu_{\ell}$ the limit of
infinite heavy-quark masses is considered, so that all the relevant
matrix elements can be expressed in terms of a single universal
function, the Isgur-Wise ($IW$) form factor \cite{SHU80,VS,IW}. The $IW$
function is explicitly calculated using our $CQ$ model wave functions of
the $B$-meson. As for the inclusive $SL$ decays, our approach is similar
to the deep inelastic scattering ($DIS$) approach by Bjorken (see Refs.
\cite{BJO_DIS}, \cite{BAR81} and \cite{JPP94}), which pictures the
heavy-meson decay as the decay of its partons. A relevant feature of
our approach is that the $SL$ width is not represented as an expansion
over the {\em small} parameter $1 / m_b$, but it is related to a
convolution of the free-quark decay tensor over the $b$-quark
distribution $|\chi(x, p_{\perp}^2)|^2$. Though this paper deals with
calculations of branching ratios, we want to point out that within our
model-dependent approach both the lepton and final hadron spectra can
be predicted in their whole kinematical accessible range. In this way
our technique can be considered complementary to model-independent
heavy-flavour methods. 

\indent The $LF$ partonic approach can be easily extended to the
description of $NL$ decays, in which the virtual $W^-$ boson
materializes into a $\bar{u} d$ (Fig. 2a) or $\bar{c} s$ (Fig. 2b)
pairs; the produced quark pair becomes one of the final hadrons, while
the $c$-quark couples with the spectator light antiquark to form other
hadrons. However, the spectator diagram is modified by the exchange of
hard gluons between initial and final quark lines. Indeed, $NL$
heavy-flavour decays are described by the effective Lagrangian of Ref.
\cite{VZS}, which contains both colour-singlet and colour-octect
four-fermion operators mixing under $QCD$ renormalization group
equations. We will refer to the contributions of these two kind of
operators as the {\em external} and {\em internal} $NL$ transitions,
respectively, the latter being characterized by a different set of quark
pairing in the final hadrons, as it is depicted in Fig. 3. The $NL$
branching ratios are estimated both adopting the so-called factorization
approximation for the matrix elements of the weak effective Lagrangian
and neglecting the perturbative $QCD$ corrections specific for each
given channel. Only the corrections due to hard-gluon exchange, yielding
the effective Lagrangian of Ref. \cite{VZS}, are taken into account (see
also \cite{ACCMM,AP91,GW79}). We stress again that in our approach both
$SL$ and $NL$ $B$-meson decays are described in terms of the same
(model-dependent) bound-state wave function of the $B$-meson, $\chi(x,
p^2_{\perp})$.

\indent The plan of the paper is as follows. In Section 2 the $CQ$ models
adopted for the description of the $B$-meson wave function are briefly
presented and the $IW$ form factor is calculated. Section 3 contains the
application of our approach to the evaluation of both exclusive and
inclusive $SL$ $B$-meson decays. In Section 4 our treatment of
{\em external} and {\em internal} $NL$ decays is presented. All the
results for the branching ratios, calculated using our $CQ$ models of
the $B$-meson wave function, are reported in Section 5. It is shown that
an overall reproduction of the available data can be achieved and the sum
of all the calculated branching ratios turns out to be very close to
unity. Moreover, we found that: ~ i) a remarkable fraction ($\sim 35
\%$) of $SL$ decay modes occur in non-$D$, non-$D^*$ final states; ~ ii)
non-perturbative effects enhance the inclusive $b \to c \bar{u}d$ decay
channels, with a sizable contribution provided by decays into heavy
mesons and baryon-antibaryon pairs; the resulting reduction of the $SL$
branching ratio brings the theoretical prediction in agreement with the
experimental value without increasing at the same time the charm
counting (i.e., the number of charm quarks produced per $b$-quark decay).
Finally, Section 6 contains our conclusions.

\section{The Isgur-Wise function.}

\indent In the limit of infinite heavy-quark masses all the relevant
matrix elements of the flavour-changing vector and axial-vector
currents, $V_{\mu} = (\bar{c} \gamma_{\mu} b)$ and $A_{\mu} = (\bar{c}
\gamma_{\mu} \gamma_5 b)$, are related to a single form factor, the $IW$
function $\xi(\eta)$, which depends only on the product $\eta$ of the
four-velocities of the initial and final hadrons. The $HQET$ allows to
write the matrix elements for pseudoscalar ($PS$) to $PS$ and $PS$ to
vector ($V$) meson transitions as follows (cf. \cite{HQET})
 \be
    <D| V_{\mu} |B> = \sqrt{M_B M_D} ~ \xi(\eta) ~ (v + v_D)_{\mu}
    \label{2.1}
 \ee
 \be
    <D^*, \lambda| V_{\mu} |B> = \sqrt{M_B M_{D^*}} ~ \xi(\eta) ~
    \varepsilon_{\mu \nu \rho \sigma} ~ v^{\rho} ~ v_{D^*}^{\sigma} ~
    e^{\nu *}(\lambda)
    \label{2.2}
 \ee
 \be
    <D^*, \lambda| A_{\mu} |B> = \sqrt{M_B M_{D^*}} ~ \xi(\eta) ~
    [e_{\mu}(\lambda) (1 + \eta) - v_{D* \mu} (v \cdot e(\lambda))]
    \label{2.3}
 \ee
where $M_B$ ($M_j$) and $v$ ($v_j$) are the mass and four-velocity of
the initial (final) heavy meson ($j = D, D^*$) and $e(\lambda)$ is the
polarization four-vector of the final vector meson with helicity
$\lambda$. As is known, the dimensionless variable $\eta \equiv v \cdot
v_j$ is related to the four-momentum transfer $q = P_B - P_j = M_B v -
M_j v_j$ by $\eta = (M_B^2 + M_j^2 - q^2) / 2 M_B M_j =$ $(1 + \zeta_j^2
- q^2 / M_B^2) / 2 \zeta_j$, where $\zeta_j \equiv M_j / M_B$. In the
limit $m_{b,c} \to \infty$ the normalization of $\xi(\eta)$ is $\xi(1) =
1$ and the leading non-perturbative corrections are quadratic in
$1/m_{b,c}$ thanks to the Luke's theorem \cite{LUK90}, which is the
analog of the Ademollo-Gatto theorem \cite{AG64} in case of the
heavy-quark symmetry.

\indent In Ref. \cite{DK93} the $IW$ function has been calculated in
the infinite momentum frame ($IMF$). Choosing the $z$ axis along the
direction of the Lorentz boost, one defines the invariant $1 - \gamma
\equiv q^+ / P_B^+ = (q_0 + q_z) / (P_{B0} + P_{Bz})$; it can be easily
verified that $q^2 = (1 - \gamma) \left( M_B^2 - M_j^2 / \gamma \right)$
and $\eta = {1 \over 2} \left( {\gamma \over \zeta_j} + {\zeta_j \over
\gamma} \right)$. Then, in terms of the $IMF$ variables $x$ and
$p_{\perp}^2$ (where $x$ is the fraction of the longitudinal $B$-meson
momentum carried by the $b$-quark and $p_{\perp}^2$ its transverse
momentum squared) the explicit expression for $\xi(\eta)$ is \cite{DK93}
 \be
    \xi(\eta) = \int d\vec{p}_{\perp} \int_0^{\gamma} dx_{sp} ~ 
    \chi(1 - x_{sp}, p_{\perp}^2) ~ \chi(1 - {x_{sp} \over
    \gamma}, p_{\perp}^2) ~ \left[1 + {(v + v_j) \cdot v_{sp} \over
    1 + \eta} \right]
    \label{2.7}
 \ee
where $x_{sp} \equiv 1 - x$, $v_{sp}$ is the four-velocity of the
spectator quark and $\chi(x, p_{\perp}^2)$ is the heavy-meson wave
function. In refs. \cite{DK93,MT96} a simple exponential ans\"{a}tz has
been considered, which in terms of the $IMF$ variables $x$ and
$p_{\perp}^2$ reads as
 \be
    \chi_{ph}(x, p_{\perp}^2) = {{\cal{N}} \over \sqrt{1 - x}} ~ \exp
    \left[-{\lambda_0 \over 2} \left( {1 - x \over \xi_0} + {\xi_0 \over
    1 - x} + {\xi_0 p_{\perp}^2 \over (1 - x) m_{sp}^2} \right) \right]
    \label{2.11}
 \ee
where $\xi_0 \equiv m_{sp}/M_B$, $m_{sp}$ is the mass of the light
spectator-quark, $\lambda_0$ is an adjustable parameter and ${\cal{N}}$
is a normalization constant fixed by the condition $\int d\vec{p}_{\perp}
dx  ~ |\chi(x, p_{\perp}^2)|^2 = 1$. Using Eq. (\ref{2.11}) in Eq.
(\ref{2.7}), the $IW$ form factor is explicitly given by (cf.
\cite{DK93})
 \be
    \xi_{ph}(\eta) = {2 \over r} {K_1(\lambda_0 r) + {2 \over r} 
    K_2(\lambda_0 r) \over K_1(2 \lambda_0) + K_2(2 \lambda_0)}
    \label{2.12}
 \ee
where $r = \sqrt{2(1 + \eta)}$ and $K_{1,2}$ are modified Bessel
functions.

\indent A relativistic approach to the construction of $CQ$ model wave
functions of mesons is provided by the Hamiltonian $LF$ formalism
\cite{LF}, where the intrinsic wave function, satisfying the correct
transformation properties under $LF$ boosts, has the following structure
(see, e.g., Refs. \cite{CAR,SIM96,DGNS96})
 \be
    \Psi_{LF}(\vec{p}_{\perp}, x; \nu \bar{\nu}) = \sqrt{{M_0 \over
    4x(1-x)} ~ \left[ 1 - \left( {m_b^2 - m_{sp}^2 \over M_0^2}
    \right)^2 \right]} ~ {w(p^2) \over \sqrt{4\pi}} ~
    {\cal{R}}(\vec{p}_{\perp}, x; \nu, \bar{\nu})
    \label{2.13}
 \ee
where $m_b$ is the mass of the $b$-quark, $M_0$ is the free mass operator
$M_0^2 = (p^2_{\perp} + m^2_b) / x + (p_{\perp}^2 + m_{sp}^2) / (1 - x)$
and the momentum dependent quantity ${\cal{R}}(\vec{p}_{\perp}, x; \nu
\bar{\nu})$ arises from the relativistic composition of the quark-spin
wave functions, with $\nu, \bar{\nu} = \pm 1/2$ being the spin
projection variables of the quarks. In Eq. (\ref{2.13}) $p^2 \equiv
p_{\perp}^2 + p_n^2$, where $p_n$ is the longitudinal momentum defined
as $p_n = (x - {1 \over 2}) M_0 + (m_{sp}^2 - m_b^2) / 2M_0$ and the
free-mass $M_0$ acquires the familiar form $M_0 = \sqrt{m_b^2 + p^2} +
\sqrt{m_{sp}^2 + p^2}$.

\indent The $LF$ wave function (\ref{2.13}) is eigenfunction of the
mass operator ${\cal{M}} = M_0 ~ + ~ {\cal{V}}$, where ${\cal{V}}$ is a
Poincar\'e-invariant interaction term. As explained in Refs.
\cite{CAR,SIM96,DGNS96}, the transformed mass operator ${\cal{R}}
{\cal{MR}}^{\dag}$ can be identified with any semi-relativistic
effective $q \bar{q}$ Hamiltonian able to reproduce meson mass spectra.
In this paper, for the latter we will consider two choices. In the first
one, the radial wave function $w(p^2)$, appearing in Eq. (\ref{2.13}),
is the solution of the following wave equation
 \be
    \left [\sqrt{m_b^2 + p^2} + \sqrt{m_{sp}^2 + p^2} + V_{GI} \right ] 
    ~ w(p^2) | 0 0 \rangle  =  M_{PS} ~ w(p^2) | 0 0 \rangle
    \label{2.17}
 \ee
where $M_{PS}$ is the mass of the $PS$ meson, $| 0 0 \rangle = \sum_{\nu 
\bar{\nu}} ~ \langle {1 \over 2} \nu {1 \over 2} \bar{\nu} |0 0 \rangle
|{1 \over 2} \nu \rangle | {1 \over 2} \bar{\nu} \rangle$ is the
canonical quark-spin wave function, and $V_{GI}$ is the effective
potential elaborated by Godfrey and Isgur ($GI$) in Ref. \cite{GI85}. The
second choice is represented by the non-relativistic ($NR$) Hamiltonian
of Ref. \cite{NCS92}, viz.
 \be
    \left [ m_b + m_{sp} + {p^2 \over 2\mu} + V_{NR} \right ] ~ w(p^2) |
    0 0 \rangle =  M_{PS} ~ w(p^2) | 0 0 \rangle
    \label{2.19}
 \ee
where $\mu \equiv m_{sp} m_b / (m_{sp} + m_b)$. Note that both the
relativized $V_{GI}$ and the non-relativistic $V_{NR}$ effective
interactions are composed by a linear-confining part (dominant at large
separations) and an effective one-gluon-exchange ($OGE$) term (dominant
at short separations), which is responsible for the hyperfine splitting
of meson mass spectra.

\indent Within the $LF$ formalism the function $\chi(x, p^2_{\perp})$
entering Eq. (\ref{2.7}) can be written in terms of the radial wave
function $w(p^2)$ as (cf. also Ref. \cite{DGNS96})
 \be
    \chi^{LF}(x, p^2_{\perp}) = \sqrt{{M_0 \over 4x(1-x)} ~ \left[ 1 -
    \left( {m_b^2 - m_{sp}^2 \over M_0^2} \right)^2 \right]} ~ {w(p^2)
    \over \sqrt{4\pi}} ~
    \label{2.20}
 \ee
We also define the distribution function $F(x)$ as the probability of
finding the $b$-quark carrying a fraction $x$ of the $B$-meson momentum,
viz.
 \be
    F(x) = \int d\vec{p}_{\perp} \sum_{\nu \bar{\nu}} ~
    \Psi_{LF}^{\dagger}(\vec{p}_{\perp}, x; \nu \bar{\nu}) ~
    \Psi_{LF}(\vec{p}_{\perp}, x; \nu \bar{\nu}) = \int d\vec{p}_{\perp}
    ~ |\chi^{LF}(x, p_{\perp}^2)|^2
    \label{2.21}
 \ee
with $\int_0^1 dx ~ F(x) = 1$. In what follows, we will refer to the
phenomenological ans\"{a}tz $\chi_{ph}$ (Eq. (\ref{2.11})) as case $A$
and to the $LF$ wave functions $\chi_{GI}^{LF}$ and $\chi_{NR}^{LF}$,
obtained from Eq. (\ref{2.20}) using the solutions of Eqs. (\ref{2.17})
and (\ref{2.19}), as cases $B$ and $C$, respectively. The values adopted
for the constituent quark masses are collected in Table 1.

\indent In Fig. 4 the $b$-quark distribution function $F(x)$, calculated
for the three cases $A$, $B$ and $C$, is shown. It can be seen that
quite similar results are obtained using $\chi_{ph}$ and
$\chi_{NR}^{LF}$, whereas the wave function $\chi_{GI}^{LF}$, obtained
from the relativized $GI$ interaction, yields a broader $x$-distribution
with the location of the peak shifted to little bit higher values of $x$.
Such differences are due to the larger content of high-momentum
components, generated by the $OGE$ term of the $GI$ interaction (see
Refs. \cite{CAR,SIM96}), and to the lower value of the $u$ ($d$)
constituent quark mass (see Table 1). The results obtained for the $IW$
form factor (Eq. (\ref{2.12}) in case $A$ and Eq. (\ref{2.7}) for cases
$B$ and $C$), multiplied by $|V_{bc}|= 0.0390$ \cite{NEU96}, are plotted
in Fig. 5 and compared with the experimental data of Refs.
\cite{ARGUS,CLEOII,ALEPH}. All the three model wave functions yield a
$IW$ form factor which is consistent with measurements; however, present
experimental uncertainties are too large to allow a stringent test on
the model wave function. The calculated slope of the $IW$ function,
$\rho^2 \equiv - d\xi(\eta) / d\eta|_{\eta = 1}$, is: $1.13$ (case $A$),
$1.03$ (case $B$) and $1.22$ (case $C$).

\section{Semileptonic decays.}

\indent In this section the formulae used for the calculation of both
exclusive and inclusive $SL$ widths are reported. The main approximations
involved are the use of the Heavy Quark Symmetry ($HQS$) for the
exclusive modes (see Eqs. (\ref{2.1}-\ref{2.3})) and the partonic
approximation for the inclusive modes (see Refs.
\cite{MT96,BJO_DIS,BAR81,JPP94}). Within our approach the $SL$ width is
not represented as an expansion over the {\em small} parameter $1 /
m_b$, but it related to a convolution of the free-quark decay tensor
over our model-dependent $b$-quark distribution $|\chi(x,
\vec{p}_{\perp}^2)|^2$. The same bound-state wave function is used for
both the exclusive and the inclusive channels. Moreover, as it will be
shown in the Section 4, our formulae for the $SL$ decays can be easily
adapted to the calculation of $NL$ modes, where non-perturbative
bound-state effects might play a different role.

\subsection{Semileptonic decays $B \to D \ell \nu_{\ell}$ and $B \to D^*
\ell \nu_{\ell}$.}

\indent The double-differential decay width for the exclusive
semileptonic process $B \to j \ell \nu_{\ell}$, where $j = D$ or $j =
D^*$, can be cast into the following form (cf. \cite{OPE,MT96})
 \be
    {d\Gamma_{j \ell \nu_{\ell}} \over dq^2 dE_{\ell}} = {G_F^2
    |V_{bc}|^2 \over 64 \pi^3} L^{\alpha \beta} \bar{W}_{\alpha
    \beta}^j 
    \label{3.1.1}
 \ee
where $q^2$ is the four-momentum squared of the dilepton system and
$E_{\ell} \equiv p_{\ell} \cdot v$ is the charged lepton energy in the
$B$-meson rest frame. In Eq. (\ref{3.1.1}) $L_{\alpha \beta}$ is the
leptonic tensor, given explicitly by
 \be
    L_{\alpha \beta}& = & {1 \over 4} ~ \sum_{spins} (\bar{\ell} ~
    \gamma_{\alpha} (1 - \gamma_5) ~ \nu_{\ell}) ~ (\bar{\nu_{\ell}} ~
    \gamma_{\beta} (1 - \gamma_5) \ell) 
    \nonumber \\
    & = & 2 \left[ p_{\ell \alpha} ~ p_{\nu_{\ell} \beta} + p_{\nu_{\ell} 
    \alpha} ~ p_{\ell \beta} - g_{\alpha \beta} ~ (p_{\ell} \cdot
    p_{\nu_{\ell}}) - i \varepsilon_{\alpha \beta \gamma \delta} ~
    p_{\ell}^{\gamma} ~ p_{\nu_{\ell}}^{\delta} \right]
    \label{3.1.2}
 \ee
and $\bar{W}_{\alpha \beta}^j$ is the (reduced) hadronic tensor, which
involves five structure functions, $\bar{W}_1^j$ to $\bar{W}_5^j$, viz.
 \be
    \bar{W}_{\alpha \beta}^j & = & {1 \over M_B^2} \sum_{\lambda} ~ (j,
    \lambda | J^{(h)}_{\alpha} |B) ~ (j, \lambda | J^{(h)}_{\beta} |B)^*
    \nonumber \\
    & = & - g_{\alpha \beta} ~ \bar{W}_1^j + v_{\alpha} ~ v_{\beta} ~
    \bar{W}_2^j -i \varepsilon_{\alpha \beta \gamma \delta} ~ v^{\gamma}
    ~ u^{\delta} ~ \bar{W}_3^j + \nonumber \\ 
    & & (v_{\alpha} ~ u_{\beta} + u_{\alpha} ~ v_{\beta}) ~ \bar{W}_4^j 
    + u_{\alpha} ~ u_{\beta} ~ \bar{W}_5^j
    \label{3.1.3}
 \ee
where $u_{\alpha} \equiv q_{\alpha} / M_B$ and $J^{(h)}_{\alpha}$ is the
charged weak hadronic current. Adopting the $HQS$ limit for the hadronic
matrix elements $(j, \lambda | J^{(h)}_{\alpha} |B)$ (see Eqs.
(\ref{2.1}-\ref{2.3})), one gets
 \be
    \bar{W}_1^D = 0 ~~~~ & , & ~~~~ \bar{W}_1^{D^*} = 2 \zeta_{D^*} 
    \eta ~ (1 + \eta) ~ \xi^2(\eta) \nonumber \\
    \bar{W}_2^D = {(1 + \zeta_D)^2 \over \zeta_D} ~ \xi^2(\eta) ~~~~ 
    & , & ~~~~
    \bar{W}_2^{D^*} = {4 \zeta_{D^*} \eta - (1 - \zeta_{D^*})^2 \over
    \zeta_{D^*}} ~ \xi^2(\eta) \nonumber \\
    \bar{W}_3^D = 0 ~~~~ & , & ~~~~ W_3^{D^*} = 2 (1 + \eta) ~ 
    \xi^2(\eta) \nonumber \\
    \bar{W}_4^D = - {(1 + \zeta_D) \over \zeta_D} ~ \xi^2(\eta) ~~~~ 
    & , & ~~~~
    \bar{W}_4^{D^*} = {(1 - \zeta_{D^*} - 2 \zeta_{D^*} \eta ) \over
    \zeta_{D^*}} ~ \xi^2(\eta)
    \nonumber \\
    \bar{W}_5^D = {1 \over \zeta_D} ~ \xi^2(\eta) ~~~~ & , & ~~~~
    \bar{W}_5^{D^*} = - {1 \over \zeta_{D^*}} ~ \xi^2(\eta)  
    \label{3.1.4}
 \ee
The integration over the lepton energy $E_{\ell}$ as well as the
remaining trace of the leptonic (\ref{3.1.2}) and hadronic (\ref{3.1.3})
tensors can be easily performed in the general case of non-vanishing
lepton masses; the final result for the exclusive $SL$ decay rate, $B \to
j \ell \nu_{\ell}$, is
 \be
    {d\Gamma_{j \ell \nu_{\ell}} \over d\eta} = {4 \over 3} ~ \Gamma_0 ~
    \Phi_{\ell} ~ \zeta_j^3 ~ (1 + \eta) ~ (\eta^2 -1 )^{1/2} ~
    F^{(j)}(\eta) ~ \xi^2(\eta)
    \label{3.1.6}
 \ee
where $\Phi_{\ell} = \sqrt{1 - 2\lambda_+ + \lambda_-^2}$, $\lambda_{\pm}
= (m_{\ell}^2  \pm m_{\nu_{\ell}}^2) / q^2$, with $m_{\ell}$ and
$m_{\nu_{\ell}}$ being the lepton masses, and
 \be
    F^{(D)}(\eta) & = & (1 + \zeta_D)^2 ~ (\eta - 1) ~ (1 + \lambda_1) ~ 
    + 3 \lambda_2 ~ (1 + \zeta^2_D - 2 ~ \zeta_D ~ \eta)
    \nonumber \\
    F^{(D^*)}(\eta) & = & \left[(1 - \zeta_{D^*})^2 ~ (1 + 5\eta) - 8 ~
    \zeta_{D^*} ~ \eta ~ (\eta - 1) \right] ~ (1 + \lambda_1) - 
    \nonumber \\
    & & 3 \lambda_2 ~ (1 + 2\eta) ~ (1 + \zeta^2_{D^*} - 2 ~ \zeta_{D^*}
    ~ \eta)
    \label{3.1.7}
 \ee
where $\Gamma_0 \equiv G_F^2 ~ |V_{bc}|^2 ~ M_B^5 / 64 \pi^3$,
$\lambda_1 = \lambda_+ - 2\lambda_-^2$ and $\lambda_2 = \lambda_+ -
\lambda_-^2$. In case of vanishing lepton masses ($m_{\ell} =
m_{\nu_{\ell}} = 0$) Eqs. (\ref{3.1.6}-\ref{3.1.7}) simplify to the well
known result \cite{ISGW}
 \be
    {d\Gamma_{D \ell \nu_{\ell}} \over d\eta} & = & {4 \over 3} ~
    \Gamma_0 ~ \zeta_D^3 ~ (1 + \zeta_D)^2 ~ (\eta^2 - 1)^{3/2} ~
    \xi^2(\eta) \nonumber \\
    {d\Gamma_{D^* \ell \nu_{\ell}} \over d\eta} & = & {4 \over 3} ~
    \Gamma_0 ~ \zeta_{D^*}^3 ~ (\eta^2 - 1)^{1/2} ~ (1 + \eta) \cdot 
    \nonumber \\
    & & \left[ (1 - \zeta_{D^*})^2 ~ (1 + 5\eta) - 8 \zeta_{D^*} ~ \eta ~
    (\eta - 1) \right] ~ \xi^2(\eta)
    \label{3.1.8}
 \ee
showing that the decay rate for the $B \to D \ell \nu_{\ell}$ transition
is suppressed by an additional kinematical factor $\eta - 1$ with
respect to the decay into the $D^*$ meson.

\subsection{Inclusive semileptonic decays.}

\indent In this subsection our technique adopted for the calculation of
inclusive $SL$ decay processes is presented. Its main feature is the
description of the decay rate in terms of the same $b$-quark distribution
function $\chi(x, p_{\perp}^2)$, introduced in the previous section,
without any explicit reference to a $1 / m_b$ expansion. Our approach to
inclusive $SL$ transitions can be considered quite analogous to the
Bjorken $DIS$ approach (see Refs. \cite{BJO_DIS,BAR81,JPP94}). As is
well known (cf. \cite{OPE}), the triple-differential decay width for the
inclusive $B \to X_c \ell \nu_{\ell}$ process is given by 
 \be
    {d\Gamma_{X_c \ell \nu_{\ell}} \over dq^2 dM_X^2 dE_{\ell}} = {G_F^2
    |V_{bc}|^2 \over 16 \pi^3} ~ {L^{\alpha \beta} ~ W_{\alpha \beta}
    \over 2M_B}
    \label{3.2.1}
 \ee
where $M_X$ is the invariant mass of the hadrons produced in the lower
block of the diagram of Fig. 1 and $W_{\alpha \beta}$ is the inclusive
hadronic tensor
 \be
    W_{\alpha \beta} = {(2 \pi)^3 \over 2M_B} \sum_{spins} ~ \sum_n ~
    \int d\tau_n ~ (n| J^{(h)}_{\alpha} |B) ~ (n| J^{(h)}_{\beta} |B)
    \label{3.2.2}
 \ee
with $d\tau_n$ being the $n$-hadron phase space. The tensor $W_{\alpha 
\beta}$ has a covariant decomposition analogous to the one of Eq.
(\ref{3.1.3}), viz.
 \be
    W_{\alpha \beta} & = & - g_{\alpha \beta} ~ W_1(t, s) + v_{\alpha}
    ~ v_{\beta} ~ W_2(t, s) -i \varepsilon_{\alpha \beta \gamma \delta}
    ~ v^{\gamma} ~ u^{\delta} ~ W_3(t, s) + \nonumber \\ 
    & & (v_{\alpha} ~ u_{\beta} + u_{\alpha} ~ v_{\beta}) ~ W_4(t, s) 
    + u_{\alpha} ~ u_{\beta} ~ W_5(t, s)
    \label{3.2.3}
 \ee
where the invariant (dimensionless) variables $t \equiv q^2 / M_B^2$ and
$s \equiv M_X^2 / M_B^2$ have been introduced. After integration over
the lepton energy $E_{\ell}$ and considering non-vanishing lepton
masses, the differential decay rate (\ref{3.2.1}) becomes
 \be
    {d\Gamma_{X_c \ell \nu_{\ell}} \over dt ds} & = & \Gamma_0 ~ M_B ~
    \Phi_{\ell}(t) ~ a(t, s) ~ \left \{ (1 + \lambda_1) ~ [2t ~ W_1(t, s)
    + {1 \over 6} a^2(t, s) ~ W_2(t, s)] + \right. \nonumber \\
    & & \left. + t \lambda_2 ~ [-4W_1(t, s) + W_2(t, s) + (1 + t - s) ~
    W_4(t, s) + t ~ W_5(t, s)] \right \}
    \label{3.2.4}
 \ee
where $a(t, s) \equiv 2|\vec{q}| / M_B = \sqrt{(1 + t - s)^2 - 4t}$.

\indent To proceed further, we consider that the $B$-meson is a bound
state of the decaying $b$-quark and a light (anti)quark-spectator
$\bar{q}_{sp}$ ($\bar{u}$ or $\bar{d}$). In the spirit of the partonic
approach the hadronic tensor $W_{\alpha \beta}$ can be written as
\cite{MT96}
 \be
    W_{\alpha \beta} = \int dx d\vec{p}_{\perp} ~ |\chi(x,
    p_{\perp}^2)|^2 ~ {1 \over x} ~ L_{\alpha \beta}^{(bc)}(p_c, p_b) ~  
    \delta \left[ (p_b - q)^2 - m_c^2 \right] ~ \Theta(E_c) 
    \label{3.2.5}
 \ee
where the tensor $L_{\alpha\beta}^{(bc)}$, describing the $b \to c ~
W^-$ transition, is defined analogously to the lepton tensor $L_{\alpha
\beta}$ of Eq. (\ref{3.1.2}). In Eq. (\ref{3.2.5}) the $\delta$-function
corresponds to the decay of a $b$-quark with longitudinal momentum $x
P_B$ and transverse momentum $\vec{p}_{\perp}$ to a $c$-quark, and
yields two roots in $x$, viz.
 \be
    \delta[(p_b - q)^2 - m_c^2] = {1 \over M_B^2 ~ (x_+ - x_-)} ~
    \left[ \delta(x - x_+) + \delta(x - x_-) \right]
    \label{3.2.6}
 \ee
with
 \be
    x_{\pm} = {1 \over 2} \left(1 + t - s \pm \sqrt{a^2(t, s) + 4 {m_c^2
    + p_{\perp}^2 \over M_B^2}} \right)
    \label{3.2.7}
\ee
Moreover, in Eq. (\ref{3.2.5}) the function $\Theta(E_c)$ has been
inserted for consistency with the use of a $CQ$ model wave function
$\chi(x, p_{\perp}^2)$ for the $b$-quark distribution in the $B$-meson.
As a matter of fact (cf. Ref. \cite{JPP94}), the root $x_-$ is related to
the contribution of the so-called $Z$-graph, arising from the negative
energy components of the $c$-quark propagator. The integration over $x$
leads to
 \be
    W_1 & = & {1 \over M_B} \int d\vec{p}_{\perp} ~ |\chi(x_+,
    p_{\perp}^2)|^2 \nonumber \\
    W_2 & = & {4 \over M_B} \int d\vec{p}_{\perp} ~ |\chi(x_+,
    p_{\perp}^2)|^2 ~ {x_+ \over x_+ - x_-} \nonumber \\
    W_3 & = & W_4 = - {2 \over M_B} \int d\vec{p}_{\perp} ~ |\chi(x_+,
    p_{\perp}^2)|^2 ~ {1 \over x_+ - x_-} \nonumber \\
    W_5 & = & 0
    \label{3.2.8}
 \ee
where we have neglected the effects of the transverse momentum in the
quark tensor $L_{\alpha \beta}^{(bc)}$. After substituting $W_i(t, s)$
from Eq. (\ref{3.2.8}) into Eq. (\ref{3.2.4}), the inclusive $SL$ decay
rate is
 \be
    \Gamma_{X_c \ell \nu_{\ell}} & = & {2 \over 3} ~ \Gamma_0 ~
    \int_{t_{min}}^{t_{max}} dt ~ \Phi_{\ell}(t) ~
    \int_{s_{min}}^{s_{max}} ds ~ a(t, s) ~ \int d\vec{p}_{\perp} ~
    |\chi(x_+, p_{\perp}^2)|^2 \cdot \nonumber \\
    & & [(1 + \lambda_1) ~ {x_+ ~ a^2(t, s) \over x_+ - x_-} + 3t ~ (1 -
    \lambda^2_-)]
    \label{3.2.9}
\ee
The integration limits in (\ref{3.2.9}) are related to the lepton
mass $m_{\ell}$ and to the threshold value $M_{th}$ at which the hadron
continuum starts in the lower block of the diagram of Fig. 1, viz.
 \be
    t_{min} = {m_{\ell}^2 \over M_B^2} ~~~~ & , & ~~~~
    t_{max} = (1 - M_{th} / M_B)^2 \nonumber \\
    s_{min} = {M_{th}^2 \over M_B^2} ~~~~ & , & ~~~~
    s_{max} = (1 - \sqrt{t})^2
    \label{3.2.10}
 \ee
The values of $M_{th}$ used in our calculations will be specified in
Section 5.

\section{Non-leptonic decay modes.}

\indent In contrast to $SL$ transitions, $NL$ processes are complicated
by the quark rearrangement mechanism due to the exchange of both soft
and hard gluons. The basic assumption in our calculation of the $NL$
amplitudes is that it is possible to separate the main contribution of
the soft gluons by incorporating all the long-distance $QCD$ effects in
the non-perturbative $CQ$ bound-state wave functions. It is well known
\cite{VZS,ACCMM,AP91,GW79} that hard-gluon exchange modifies the weak
forces driving $NL$ heavy-flavour decays; at the mass scale $\mu = M_W$
the weak Lagrangian is
 \be
    L_W(\mu = M_W) = - {4G_F \over \sqrt{2}} \left[ V_{cb} ~ (\bar{c}_L
    \gamma_{\mu} b_L) + V_{ub} ~ (\bar{u}_L \gamma_{\mu} b_L) \right]
    \cdot \left[ V^*_{ud} ~ (\bar{d}_L\gamma_{\mu} u_L) + V^*_{cs} ~
    (\bar{s}_L \gamma_{\mu} c_L) \right]
    \label{4.0.1}
 \ee
where we have ignored the $b \to t$ coupling term. Radiative $QCD$
corrections lead to an effective Lagrangian at the mass scale $\mu =
m_b$ \cite{VZS}
 \be
    L^{eff}_W(\mu = m_b) & = & - {4G_F \over \sqrt{2}} ~ V_{cb} V^*_{ud}
    \left\{C_1 (\bar{c}_L \gamma_{\mu} b_L) (\bar{d}_L \gamma_{\mu} u_L)
    + C_2 (\bar{d}_L \gamma_{\mu} b_L) (\bar{c}_L \gamma_{\mu} u_L) +
    \right. \nonumber \\ & & \left.
    (\bar{d},u) \leftrightarrow (\bar{s},c) \right\}
    \label{4.0.2}
 \ee
which contains both a colour-singlet, $O_1 \equiv (\bar{c}_L
\gamma_{\mu} b_L) (\bar{d}_L \gamma_{\mu} u_L)$, and a colour-octet,
$O_2 \equiv (\bar{d}_L \gamma_{\mu} b_L) (\bar{c}_L \gamma_{\mu} u_L)$,
four-fermion transition operators. The factors $C_1 \pm C_2 = C_{\pm}$
are the Wilson coefficients determined by renormalization group
equations, which allows to move from the mass scale $\mu = M_W$ to the
lower scale $\mu = m_b$. In the leading-log approximation one has
$C_{\pm} = L^{\gamma_{\pm}}$, where $L = \ln(M_W / \Lambda_{QCD}) /
\ln(m_b / \Lambda_{QCD})$, with $\Lambda_{QCD}$ being the $QCD$ scale,
and $\gamma_+ = 6 / (33 - 2N_f) = - {\gamma_- \over 2}$. Including
next-to-leading logarithmic corrections one obtains $C_1 =1.13$ and
$C_2 = -0.29$ \cite{VZS,GW79}, showing explicitly that the Wilson
coefficient $C_2$ is colour-suppressed (i.e., $C_2 \sim 1 / N_c$, where
$N_c$ is the number of colours). In Eq. (\ref{4.0.2}) the contributions
of the so-called $W$-exchange and weak annihilation diagrams have been
neglected because they are not relevant for the transitions considered
(cf. \cite{BGR86}).

\indent In the next two subsections we will discuss separately the {\em
external} $NL$ decays, proceeding via the colour-singlet operator
$O_1$ (Fig. 2), and the {\em internal} $NL$ decays related to the
colour-suppressed transition operator $O_2$ (Fig. 3). The operators
$O_1$ and $O_2$ mix each other under $QCD$ renormalization group
equations and therefore the separation among the {\em external} and {\em
internal} diagrams is scale dependent. We point out that internal decays
of the $\bar{B}^0-$meson lead to final hadron states different from
those corresponding to external decays, namely, the external decays of
the $\bar{B}^0$ meson due to $W^- \to \bar{c} s$ transitions lead to
$(\bar{c} s) + (\bar{d} c)$ final states (like, e.g., $D_s^{*-} D^{*+}$),
while in case of internal decays $(\bar{d} s) + (\bar{c} c)$ states
(like, e.g., $K^{*0} + J/\psi$) are produced. The same is valid for $W^-
\to \bar{u} d$ transitions, where the internal decays lead to $(\bar{d}
d) + (\bar{u} c) $ hadron states (like, e.g., $\rho^0 + D^{*0}$), whereas
the external ones lead to $(\bar{u} d) + (\bar{d} c)$ states (like,
e.g., $\rho^- + D^{*+}$)\footnote{In case of the $B^+$ meson the
internal decays due to $W^- \to \bar{u} d$ transitions leads to exactly
the same states as the external decays. Nevertheless, we can safely
disregard the interference term between the external and internal
amplitudes, which is expected to be very small because the total decay
widths of the $\bar{B}^0$ and $B^+$ mesons coincide with a good
accuracy.}.

\subsection{{\em External} non-leptonic decays.}

\indent First, we discuss the amplitudes of the transitions to the
continuum states $X_{\bar{u}d}$ and $X_{\bar{c}s}$ in the upper blocks
of Fig. 2. The corresponding decay amplitudes are obtained from the
first term in $L^{eff}_W$ (Eq. (\ref{4.0.2})) in a way quite similar to
the $SL$ case. We use the zero-order approximation in the $1/N_c$
expansion and neglect the radiative $QCD$ corrections different from
those already included in the Wilson coefficients $C_{1,2}$. In this
approximation all the amplitudes for the decays $\bar{B}^0 \to
X_{\bar{u}d} D^+ (D^{*+})$, $\bar{B}^0 \to X_{\bar{c}s} D^+ (D^{*+})$,
$\bar{B}^0 \to X_{\bar{u}d} ~ X_{\bar{d}c}$ and $\bar{B}^0 \to
X_{\bar{c}s} ~ X_{\bar{d}c}$ are proportional to the $NL$ enhancement
factor $C_1$. We disregard the contribution of the second term in Eq.
(\ref{4.0.2}), because it is proportional to the colour-suppressed
coefficient $C_2$. With these simplifications the matrix elements of
interest are of the same type as the ones found in the analysis of $SL$
decays in Sec. 3. The final result reduces to replace in Eqs.
(\ref{3.1.6}) and (\ref{3.2.9}) the factor $\Gamma_0$ with $N_c ~ C_1^2
~ \Gamma_0$ and also to consider in Eq. (\ref{3.1.2}) the $CQ$ masses
instead of the lepton masses.

\indent In case of the production of a single meson (e.g. $D_s$, $D^*_s$,
...) in the upper block of Fig. 2 the corresponding amplitudes can be
obtained from Eq. (\ref{3.1.1}) by substituting the leptonic current
$J^{\ell \nu_{\ell}}_{\alpha} = \bar{\ell} \gamma_{\alpha} (1 -
\gamma_5) \nu_{\ell}$ with $J_{\alpha}^{PS} = f_{PS} ~ q_{\alpha}$ and $
J_{\alpha}^V = f_V ~ M_V ~ e_{\alpha}$, where $f_{PS}$ and $f_V$ are the
coupling constants of $PS$ and $V$ mesons to the $W$-boson,
respectively, and $q_{\alpha}$ is the $W$-boson four-momentum. Then,
instead of the leptonic tensor $L_{\alpha \beta}$ of Eq. (\ref{3.1.2})
one has
 \be
    {1 \over 4} \sum_{spins} ~ j^i_{\alpha} ~ j^i_{\beta} = \left\{
    \ba{ll} {1 \over 4} M_{PS}^2 ~ f_P^2 ~ u_{\alpha} ~ u_{\beta} 
    \hfill \qquad \qquad i = PS \\ [3mm]
    {1 \over 4} M_V^2 ~ f_V^2 ~ (u_{\alpha} ~ u_{\beta} - g_{\alpha
    \beta}) \hfill \qquad \qquad i = V \ea \right.
    \label{4.1.2}
 \ee
Denoting by $\Gamma_{ij}$ the partial $NL$ width, where the index $i$
refers to a $PS$ or $V$ meson produced in the upper blocks of the
diagrams of Fig. 2 and the index $j$ denotes a charmed, resonant or
non-resonant, hadron state ($D, D^*, X_c$) produced in the lower blocks,
one gets (cf. also \cite{MT96})
 \be
    \Gamma_{ij} / \Gamma_0 & = & {\pi^2 C_1^2 f_i^2 \over 2M_B^2} ~ [(1 +
    \zeta_j)^2 - \zeta_i^2]^2 ~ (\zeta_j^{-1/2} - \zeta_j^{1/2})^2 ~
    a(\zeta_i^2, \zeta_j^2) ~ \xi^2(\eta) \hfill \qquad i = j = PS
    \nonumber \\
    \Gamma_{ij} / \Gamma_0 & = & {\pi^2 C_1^2 f_i^2 \over 2M_B^2} ~
    (\zeta_j^{-1/2} + \zeta_j^{1/2})^2 ~ a^3(\zeta_i^2, \zeta_j^2) ~
    \xi^2(\eta) \hfill \qquad \qquad i, j = PS, V ~ \mbox{or} ~ V, PS
    \nonumber   \\
    \Gamma_{ij} / \Gamma_0 & = & {\pi^2 C_1^2 f_i^2 \over 2M_B^2} ~ [(1 +
    \zeta_j)^2 - \zeta_i^2] ~ (R_{ij}/\zeta_j) ~ a(\zeta_i^2, \zeta_j^2)
    ~ \xi^2(\eta) \hfill \quad \qquad \qquad i = j = V
    \nonumber \\
    \Gamma_{ij} / \Gamma_0 & = & {4 \pi^2 C_1^2 f_i^2 \over M_B^2} \int
    \limits_{s_{th}}^{(1 - \zeta_i)^2} ds d\vec{p}_{\perp} ~
    |\chi(\tilde{x}_+, p^2_{\perp})|^2 ~ {a(\zeta_i^2, s) \over
    \tilde{x}_+ - \tilde{x}_-} \cdot \nonumber \\
    & & \left[\tilde{x}_+ (\tilde{x}_+ + \tilde{x}_-)^2 - (\tilde{x}_- +
    3\tilde{x}_+) ~ \zeta_i^2 \right] \hfill \qquad \qquad \qquad i = PS;
    ~ j = X_c
    \nonumber\\
    \Gamma_{ij} / \Gamma_0 & = & {4 \pi^2 C_1^2 f_i^2 \over M_B^2} \int
    \limits_{s_{th}}^{(1 - \zeta_i)^2} ds d\vec{p}_{\perp} ~
    |\chi(\tilde{x}_+, p^2_{\perp})|^2 ~ {a(\zeta_i^2, s) \over
    \tilde{x}_+ - \tilde{x}_-} \cdot \nonumber \\
    & & \left[\tilde{x}_+ (\tilde{x}_+ + \tilde{x}_-)^2 - (3\tilde{x}_- +
    \tilde{x}_+) ~ \zeta_i^2 \right] \hfill \qquad \qquad \qquad i = V; 
    ~ j = X_c
    \label{4.1.3}
 \ee
where $R_{ij} = (1 - \zeta_j^2)^2 - \zeta_i^2 [(1 - \zeta_j)^2 -
8\zeta_j]$, $\tilde{x}_{\pm} = x_{\pm}(t = \zeta_i^2, s)$ and $s_{th}
\equiv M_{th}^2 / M_B^2$. The complete list of the partial widths
corresponding to {\em external} $NL$ decays can be found in Ref.
\cite{MT96}.

\subsection{{\em Internal} non-leptonic decays.}

\indent The graphs shown in Fig. 3 correspond to the so-called {\em
internal} $NL$ decays of the $B$-meson. The upper block in Fig. 3(a)
represents a hadron state formed by a $c$-quark resulting from the $b
\to c W^-$ transition and a $\bar{u}$-quark resulting from the $W^- \to
\bar{u} d$ decay. In Fig. 3(b) the $W^- \to \bar{c}s$ decay leads to the
production of a colourless $c \bar{c}$ state ($\eta_c$, $J/\psi$ or $c
\bar{c}$ in the continuum, $X_{\bar{c}c}$) in the upper block and a
strange hadron state ($K$, $K^*$, $X_{\bar{d}s}$) in the lower block.
The graphs of Fig. 3 correspond to the second term in the $VZS$
Lagrangian (\ref{4.0.2}) and yield, after neglecting $QCD$ corrections,
the same factorized expressions already found for the $SL$ decays. The
only difference is that we have to replace $C_2$ for $C_1$ in Eq.
(\ref{4.1.3}) and the factor $\Gamma_0$ with $N_c ~ C_2^2 ~ \Gamma_0$ 
in Eqs. (\ref{3.1.6}) and (\ref{3.2.9}). After these changes the rest of
the calculation is essentially the same as in the case of the {\em
external} $NL$ decays.

\indent The diagrams shown in Fig. 6(a) describe the production of a
coloured diquark $cd$ and anti-diquark ($\bar{u} \bar{d}$), while the
similar graphs in Fig. 6(b) correspond to the production of a $cs$
diquark and a $\bar{c} \bar{d}$ anti-diquark. After making a Fierz-like
transformation in Eq. (\ref{4.0.2}) (exchanging the $\bar{c}$ and $u$
quark fields) the $VZS$ Lagrangian can be written in the form
 \be
    L^{eff}_W(\mu = m_b) & = & - {4G_F \over \sqrt{2}} ~ V_{cb} ~
    V_{ud}^* \left \{ C_- (\bar{c}^c_L \gamma_{\mu}
    d_L)^{\dagger}_{\bar{3}} ~ (\bar{u}^c_L \gamma_{\mu} b_L)_3 + 
    \right. \nonumber \\ 
    & & \left. C_+ (\bar{c}^c_L \gamma_{\mu} d_L)^{\dagger}_{\bar{6}} ~
    (\bar{u}^c_L \gamma_{\mu} b_L)_6 + (d, \bar{u}) \to (s, \bar{c})
    \right \}
    \label{4.2.1}
 \ee
This form suggests that diquarks can be produced in $\bar{3}$ and $3$
colour states. The corresponding decay amplitudes are proportional to
the factor $C_- / 3 \simeq 0.47$. When flying away, diquarks can pick up
the light quark of a $q \bar{q}$ pair produced from the vacuum due to
the confinement mechanism. As a result, a colourless charmed baryon
$(cd)q$ and a colourless charmless antibaryon $(\bar{u} \bar{d})
\bar{q}$ are produced in the $B$-meson decay (with any number of final
mesons). Such a mechanism for the production of baryon-antibaryon states
has been firstly suggested in Ref. \cite{PS93}. The corresponding
diagrams are shown in Fig. 6(a), while similar graphs, corresponding to
the decay $\bar{B}^0 \to \Xi_{cs} + \bar{\Lambda}_c$, are depicted in
Fig. 6(b).

\indent Instead of calculating many possible channels corresponding to
the production of particles both in the upper and lower blocks, we have
actually calculated the decay probability for the production of a
diquark and anti-diquark with the effective mass $M = \sqrt{q^2}$ and
$M'$ larger than the mass of the corresponding baryon. The resulting
decay widths are given by Eqs. (\ref{3.2.9}-\ref{3.2.10}) with the
replacements $m_{\ell} \to M_{\Lambda_c}$ ($M_{\Xi_c}$) and $M_{th}
\to M_N$ ($M_{\Lambda_c}$), obtaining in this way our approximation for
the calculation of the inclusive probability of production of
baryon-antibaryon states with any number of final mesons.

\section{Numerical results.}

\indent In what follows we will denote by $\Gamma_{ij}$ the partial
width for the $B$-meson decays depicted in Fig. 1 for the $SL$ modes and
in Figs. 2, 3 and 6 for the $NL$ modes. The index $i$ refers to the
lepton pair $\ell \nu_{\ell}$ or to the hadron states produced in the
upper block of the diagrams ($\pi,\rho,...,$ or charmless continuum
$X_{\bar{u}d}$ for the diagram 2(a), and $D_s, D_s^*, ..., X_{\bar{c}s}$
for the diagram 2(b)), while the index $j$ denotes a charmed hadron
state ($D, D^*, X_c$) produced in the lower blocks. The partial width
can be written in the form $\Gamma_{ij} = \Gamma_0 ~ \beta_{ij}$, where
$\beta_{ij}$ is a dimensionless quantity and $\Gamma_0$ is given
explicitly by
 \be
    \Gamma_0 = {G_F^2 M_B^5 \over (4 \pi)^3} ~ |V_{bc}|^2 = (4.27 \pm
    0.44) \cdot 10^{-4} ~ eV
    \label{5.1}
 \ee
when the values $|V_{bc}| = 0.039 \pm 0.002$ \cite{NEU96} and $M_B =
(5.279 \pm 0.002) ~ GeV$ \cite{PDG96} are considered. Then, the total
calculated width is given by $\Gamma_B^{tot} = \Gamma_0 ~ \sum_{ij}
\beta_{ij}$, where the sum runs over all decay modes of the $B$-meson. In
our phenomenological approach the partial decay widths depend on the
following set of parameters:

i) the masses of the constituent quarks building up final mesons and
multi-hadrons and the masses of pseudoscalar and vector heavy mesons; the
former ones are listed in Table 1, while the latter have been taken from
$PDG$ \cite{PDG96};

ii) the coupling constants $f_{PS}$ and $f_V$ of the $PS$ and $V$ mesons
to the $W$ boson; in units of $GeV$ they are given by (cf. Ref.
\cite{MT96})
 \be
     f_{\pi} & = & 0.13 ~~~~ f_D = 0.17 ~~~~ f_{D_s} = 0.26 ~~~~
     f_{\eta_c} = 0.38 \nonumber \\
     f_{\rho} & = & 0.20 ~~~~ f_{D^*} = 0.20 ~~~~ f_{D^*_s} = 0.29 ~~~~ 
     f_{J / \psi} = 0.41
     \label{5.2}
 \ee

iii) the threshold values $M_{th}$ at which the hadron continuum starts;
in our calculations we have adopted the following values (in units of
$GeV$):
 \be
    (M_{th})_{c \bar{d}} & = & 2.10 ~~~~ (M_{th})_{c \bar{s}} = 2.26 ~~~~
    (M_{th})_{c \bar{c}} = 3.25 \nonumber \\
    (M_{th})_{d \bar{u}} & = & 1.00 ~~~~ (M_{th})_{d \bar{s}} = 1.06
    \label{5.3} 
 \ee
for decays into mesons, and
 \be
    (M_{th})_{c \bar{d}} & = & 1.96 ~~~~ (M_{th})_{c \bar{s}} = 1.65 ~~~~
    (M_{th})_{d \bar{u}} = 0.45
    \label{5.4}
 \ee
for the decays into baryon-antibaryon pair.

\indent The evaluation of the partial widths $\Gamma_{ij}$ depend upon
the form of the $b$-quark distribution inside the $B$-meson and, in case
of exclusive channels, they depend explicitly upon the $IW$ function
$\xi(\eta)$. As described in Section 2, we have used three different
models for the $B$-meson wave function $\chi(x, p_{\perp}^2)$. Two of
them, $\chi_{GI}^{LF}$ and $\chi_{NR}^{LF}$, are fixed by the choice of
the $q \bar{q}$ potential in Eqs. (\ref{2.17}) and (\ref{2.19}),
respectively. The third one is the phenomenological ans\"{a}tz of Eq.
(\ref{2.11}), where the parameters $\lambda_0 = 1.0$ and $\xi_0 = m_{sp}
/ M_B = 0.05$ have been fixed by fitting the $CLEO-II$ data \cite{CLEOII}
(see Fig. 5).

\indent Using these parameters we have calculated the partial widths
corresponding to $SL$ (exclusive and inclusive) and $NL$ (external and
internal) decay modes; our detailed results, given in terms of
$\Gamma_{ij} / \Gamma_0$, are collected in Tables 2-7. First of all,
we want to show that our theoretical estimate of the total $B$-meson
width, $\Gamma_B^{tot}$, is consistent with experimental data. Indeed,
from Tables 2-7 the sum of all the calculated partial widths
$\Gamma_{ij}$ yields: $0.981 ~ \Gamma_0 ~ (A), ~ 1.041 ~ \Gamma_0 ~ (B)$
and $1.0779 ~ \Gamma_0 ~ (C)$. Adding a contribution of $\simeq 3 \%$ due
to the charmless (direct $b \to u$ and penguin $b \to s$) transitions
\cite{AP91}, one gets: $\Gamma_B^{tot} = 1.011 ~ \Gamma_0 ~ (A), ~ 1.071
~ \Gamma_0 ~ (B)$ and $1.107 ~ \Gamma_0 ~ (C)$. At $|V_{bc}| = 0.039 \pm
0.002$ \cite{NEU96}, one has: $\Gamma_B^{tot} = (4.32 \pm 0.45) \cdot
10^{-4} ~ eV ~ (A), ~ (4.57 \pm 0.47) \cdot 10^{-4} ~ eV ~ (B)$ and
$(4.73 \pm 0.49) \cdot 10^{-4} ~ eV ~ (C)$. Our predictions compare
favourably with the experimental value $(4.19 \pm 0.11) \cdot 10^{-4}
~ eV$ obtained from the updated world-average value of the
$\bar{B}^0$-meson life-time, $\tau_{\bar{B}^0} = 1.57 \pm 0.04 ~ ps$
\cite{BHP96}.

\indent  A summary of our results for the major branching ratios,
${\cal{B}}r_{ij} \equiv \Gamma_{ij} / \Gamma_B^{tot}$, is presented in
Tables 8-9 and compared with updated world-average data. Moreover, our
results for the inclusive $B$-meson branching ratios corresponding to the
elementary transitions $b \to c \ell \nu_{\ell}$, $b \to c \bar{c}s$ and
$b \to c \bar{u}d$ are collected in Table 10. Since the net effect of
radiative $QCD$ corrections is expected to be of the same order of
magnitude of the uncertainties related to the choice of the $B$-meson
wave function, in Tables 8-10 we have reported also the average of our
predictions for cases $A$, $B$ and $C$ with an assigned error given by
the standard deviation from the mean value. It can be seen that a
remarkable overall agreement with the data, including the inclusive
branching ratio into charmed baryons and the so-called charm counting
(i.e., the number of charm quarks produced per $b$-quark decay), is
achieved. A throughout comparison of our results with those of other
approaches is out of the scope of this paper. We will limit ourselves to
the following two comments.

\begin{enumerate}

\item In the parton model \cite{AP91} the $SL$ branching ratio
${\cal{B}}r_{SL} \equiv {\cal{B}}r(B \to X_c e \nu_e)$ is $\simeq 13 \%$
and the charm counting $n_c$ is $\simeq 1.15$. The inclusion of
non-perturbative corrections through the $HQET$ expansion leads only to
${\cal{B}}r_{SL} \ga 12.5 \%$ \cite{BBSV94}. In Ref. \cite{BBBG95} it has
been shown that higher-order radiative $QCD$ corrections can increase the
partial width of the $b \to c \bar{c}s$ processes, decreasing in this
way the $SL$ branching ratio, but at the price of increasing the charm
counting ($n_c \ga 1.25$); moreover, the result for ${\cal{B}}r_{SL}$
turns out to be significantly scale-dependent \cite{NEU96}. Within our
phenomenological quark model the calculated $SL$ branching ratio
(${\cal{B}}r_{SL} \simeq 11\%$) is in nice agreement with the updated
world-average value $10.90 \pm 0.46$ \cite{NEU96}. At the same time, our
prediction for the charm counting ($n_c \simeq 1.20$) compares
favourably with recent experimental results, $n_c^{exp} = 1.16 \pm 0.05$
\cite{CLEO_C} and $n_c^{exp} = 1.23 \pm 0.07$ \cite{ALEPH_C}. Moreover,
our prediction for the inclusive branching ratio due to the elementary
$b \to c \bar{c}s$ transitions, is in accord with experimental findings
(see Table 10). Therefore, our results imply that non-perturbative
effects, modeled by our $B$-meson wave function $\chi(x, p_{\perp}^2)$,
can enhance significantly the inclusive $b \to c \bar{u}d$ decay modes,
leading to a sizable reduction of the $SL$ branching ratio without
increasing at the same time the charm counting. We point out that a
sizable fraction of such an enhancement is provided by {\em internal}
$NL$ decays into heavy mesons and baryon-antibaryon pairs (see Table
7).  

\item Our results for exclusive as well as inclusive $SL$ branching
ratios are in nice agreement with updated experimental data (see Table
8). It follows that the exclusive $SL$ decays into $D$ and $D^*$ mesons
account for $\sim 65 \%$ of the total $SL$ branching ratio, leaving a
remarkable fraction of $\sim 35 \%$ to decays into non-$D$ and non-$D^*$
channels. This is at variance with the result of the $ISGW$ model of Ref.
\cite{ISGW}, which yield only a value of $\sim 10 \%$ for the probability
of non-$D$, non-$D^*$ channels.

\end{enumerate}

\indent Before closing, we want to emphasize that the comparison among
the electron spectrum calculated within our quark model and the $ISGW$
one has been already carried out in Ref. \cite{GNTS96}, where it has also
been shown that our predicted spectrum is consistent with the results
obtained in Ref. \cite{MN94} through a partial $OPE$ resummation
performed in the end-point region.

\section{Conclusions.}

\indent A relativistic quark model has been applied to the description
of semileptonic and non-leptonic charmed decays of the $B$-meson.
Non-perturbative $QCD$ effects have been mocked up by a constituent quark
wave function for the $B$-meson, describing the motion of the $b$-quark
inside the $B$-meson. Different approximations for the latter,
either based on a phenomenological ans\"{a}tz or derived from analyses
of the meson spectra, have been adopted. In particular, two wave
functions, constructed via the Hamiltonian light-front formalism using a
relativized and a non-relativistic constituent quark model, have been
considered, obtaining a link between standard spectroscopic quark models
and the $B$-meson decay physics.

\indent As for the exclusive semileptonic decay processes $B \to D \ell
\nu_{\ell}$ and $B \to D^* \ell \nu_{\ell}$, the universal Isgur-Wise
function and the semileptonic branching ratio (as well as the lepton and
hadron distributions in the final states) can be calculated in terms of
our model $B$-meson wave function.

\indent A partonic approach has been applied to the description of
inclusive $B$-meson decays to multi-hadrons. Within our approach both the
spectra and the decay probabilities can be expressed in terms of the same
bound-state wave function used for the description of exclusive
semileptonic channels, without any explicit reference to a $1 / m_b$
expansion. The main drawback of our approach is likely to be connected
with the lack of the effects due to the quark interaction in the final
hadronic states. Indeed, our distribution of produced hadron masses has
a maximum at the threshold value, i.e., just in the region where quark
interaction might be important. Moreover, radiative $QCD$ corrections
have been neglected; however, they can be easily introduced using
standard methods. Their net effect ($\sim 10 \div 20 \%$) is expected to
be of the same order of magnitude of the uncertainties related to the
choice of the $B$-meson wave function.

\indent The calculated sum of all the major branching ratios turns out to
be close to unity. A remarkable overall agreement with updated
world-average data, including the inclusive branching ratio into charmed
baryons and the charm counting, has been achieved. In particular, we
have found that non-perturbative effects can enhance the inclusive width
corresponding to elementary $b \to c \bar{u}d$ transitions. In this
respect an important contribution is provided by {\em internal}
non-leptonic decays into heavy mesons and baryon-antibaryon pairs.
Correspondingly, the semileptonic branching ratio is brought in
agreement with its experimental value without increasing, at the same
time, the charm counting. Finally, a remarkable fraction ($\sim 35 \%$)
of semileptonic decay modes have been found to occur in non-$D$,
non-$D^*$ final states.   

\section*{Acknowledgments.}

We are indebted to P. Kulikov for having checked the results of our
calculations. Two of the authors (I.L.G. and I.M.N.) acknowledge the
financial support of the INTAS grant No. 93-0079. This work was done in
part under the RFFR grant, Ref. No. 95-02-04808a.

\newpage

\noindent {\bf Table 1}. Values of the constituent quark masses adopted
for our models of the $B$-meson wave function $\chi(x, p_{\perp}^2)$.
Case A is the phenomenological wave function of Eq. (\ref{2.11}). Cases
B and C correspond to the light-front wave functions $\chi_{GI}^{LF}$ and
$\chi_{NR}^{LF}$ (see Eq. (\ref{2.20})), derived via the $LF$ formalism
from the relativized \cite{GI85}  (Eq. (\ref{2.17})) and the
non-relativistic \cite{NCS92} (Eq. (\ref{2.19})) constituent quark
models, respectively.

\begin{center}

\begin{tabular}{||c||c|c|c|c||}
\hline
Case & $m_u$ & $m_s$ & $m_c$ & $m_b$ \\ \hline
 $A$ & 0.265 & 0.560 & 1.400 & 5.279 \\
 $B$ & 0.220 & 0.419 & 1.628 & 4.977 \\
 $C$ & 0.337 & 0.576 & 1.835 & 5.237 \\ \hline
\end{tabular}

\end{center}

\vspace{3cm}

\noindent {\bf Table 2}. Partial widths $\Gamma_{ij}$ for semileptonic
charmed decays of the $B$-meson in units of $\Gamma_0$ (see Eq.
(\ref{5.1})). Cases A, B and C correspond to the parameter sets
of Table 1. The row labeled $\Sigma$ correspond to the total
semileptonic branching ratio calculated for each final lepton pair. 

\begin{center}

\begin{tabular}{||l||lll||lll||lll||} \hline
$j \setminus i$                &
\multicolumn{3}{c||}{$e\nu$}   &
\multicolumn{3}{c||}{$\mu\nu$} &
\multicolumn{3}{c||}{$\tau\nu$}  \\ \hline
& A& B& C& A& B& C& A& B& C      \\ \cline{2-10}

$D$   & 1.83& 2.07& 1.69& 1.82& 2.06& 1.69&0.56&0.59&0.53 \\
$D^*$ & 5.54& 5.98& 5.26& 5.52& 5.96& 5.24&1.39&1.45&1.34 \\
$X_c$ & 3.10& 3.99& 5.88& 3.08& 3.97& 5.84&0.36&0.47&0.75 \\ \hline
\multicolumn{1}{l||}{$\Sigma$}
      &10.47&12.04&12.83&10.42&11.99&12.77&2.31&2.51&2.62 \\
\cline{2-10}
\end{tabular}

\end{center}

\vspace{3cm}

\noindent {\bf Table 3}. Partial widths for the non-leptonic charmed
decays of the $\bar{B}^0$-meson corresponding to the {\em external}
transition $W^- \to \bar{u} d$ (see Fig. 2(a)), in units of $\Gamma_0$.
The notations are the same as in Table 2.

\begin{center}

\begin{tabular}{||l||lll||lll||lll||} \hline
$j\setminus i$                        &
\multicolumn{3}{c||}{$\pi^-$}         &
\multicolumn{3}{c||}{$\rho^-$}        &
\multicolumn{3}{c||}{$X^-_{\bar{u}d}$}  \\ \hline
& A& B& C& A& B& C& A& B& C             \\ \cline{2-10}
$D^+$   &0.40&0.49&0.36&0.90&1.08&0.82& 5.96& 6.67& 5.34 \\
$D^{*+}$&0.42&0.49&0.38&1.08&1.25&0.99&19.43&20.87&18.51 \\
$X_c^+$ &0.75&0.92&1.26&1.73&2.15&2.96& 9.53&12.36&17.93 \\ \hline
\multicolumn{1}{l||}{$\Sigma$}
        &1.57&1.90&2.00&3.71&4.48&4.77&34.92&39.90&41.78 \\ \cline{2-10}
\end{tabular}

\end{center}

\newpage

\noindent {\bf Table 4}. The same as in Table 3, but for the {\em
external} $W^- \to \bar{c} s$ transition (see Fig. 2(b)).

\begin{center}

\begin{tabular}{||l||lll||lll||lll||} \hline
$j\setminus i$&
\multicolumn{3}{c||}{$D^-_s$}         &
\multicolumn{3}{c||}{$D^{*-}_s$}      &
\multicolumn{3}{c||}{$X^-_{\bar cs}$}   \\ \hline
& A& B& C& A& B &C& A& B& C             \\ \cline{2-10}
$D^+$      &1.55&1.75&1.13&1.28&1.43&1.19& 2.42&2.25&2.32 \\
$D^{*+}$   &1.09&1.20&1.02&3.35&3.66&3.15& 6.66&5.71&6.46 \\
$X^+_{cd}$ &1.29&1.63&2.44&2.28&3.03&4.58& 1.58&1.62&1.18 \\ \hline
\multicolumn{1}{l||}{$\Sigma$}
           &3.93&4.58&4.59&6.91&8.12&8.92&10.66&9.58&9.96 \\ \cline{2-10}
\end{tabular}

\end{center}

\vspace{2cm}

\noindent {\bf Table 5}. Partial widths for the {\em internal} $NL$
charmed decays of the $\bar{B}^0$-meson corresponding to the transition
$W^- \to \bar{u} d$ (see Fig. 3(a)), in units of $\Gamma_0$.

\begin{center}

\begin{tabular}{||l||lll||lll||lll||} \hline
&
\multicolumn{3}{c||}{$D^0$}          &
\multicolumn{3}{c||}{$D^{*0}$}       &
\multicolumn{3}{c||}{$X_{\bar{u}c}$}   \\ \hline
 & A & B & C & A & B & C & A & B & C   \\ \cline{2-10}
$\pi^0$ & $< 0.01$ & $< 0.01$ & $< 0.01$ & $< 0.01$ & $< 0.01$
& $< 0.01$ & $< 0.01$ & $< 0.01$ & $< 0.01$ \\
$\rho^0$ & 0.01 & 0.01 & 0.01 & 0.02 & 0.02 & 0.01
& 0.62 & 0.62 & 0.43 \\
$X^0_{\bar{d}d}$ & 0.15 & 0.13 & 0.16 & 0.31 & 0.28 & 0.34 & 1.40 & 0.98
& 0.98 \\ \hline
\multicolumn{1}{l||}{$\Sigma$} & 0.16 & 0.14 & 0.17 & 0.33 & 0.30 & 0.35
& 2.02 & 1.60 & 1.41 \\ \cline{2-10}
\end{tabular}

\end{center}

\vspace{2cm}

\noindent {\bf Table 6}. The same as in Table 5, but for the {\em
internal} $W^- \to \bar{c} s$ transition (see Fig. 3(b)).

\begin{center}

\begin{tabular}{||l||lll||lll||lll||} \hline
&
\multicolumn{3}{c||}{$ \eta_c $}&
\multicolumn{3}{c||}{$J / \psi$}&
\multicolumn{3}{c||}{$X_{\bar{c}c}^0$} \\ \hline
& A & B & C & A & B & C & A & B & C    \\ \cline{2-10}
$K^0$    & 0.02 & 0.02 & 0.01 & 0.01 & 0.02 & 0.01 & 0.06 & 0.06 &
0.03 \\
$K^{*0}$ & 0.05 & 0.06 & 0.05 & 0.14 & 0.15 & 0.12 & 0.39 & 0.31 &
0.12 \\
$X^0_{\bar ds}$ & 0.32 & 0.25 & 0.34 & 0.56 & 0.44 & 0.60 & 0.29 & 0.12 &
0.04 \\ \hline
\multicolumn{1}{l||}{$\Sigma$} & 0.39 & 0.33 & 0.40 & 0.71 & 0.61 &
0.73 & 0.74 & 0.49 & 0.19 \\ \cline{2-10}
\end{tabular}

\end{center}

\vspace{2cm}

\noindent {\bf Table 7}. Values of the inclusive partial widths for the
$\bar{B}^0 \to \bar{N} \Lambda_c + X$ and $\bar{B}^0 \to \Xi_{cs}
\bar{\Lambda}_c + X$ decays (see Fig. 6), in units of $\Gamma_0$.

\begin{center}

\begin{tabular}{||l|l|l||}
\multicolumn{3}{c}{$\bar{B}^0 \to \bar{N} \Lambda_c + X$} \\ \hline
   A &    B &    C \\ \hline
6.80 & 4.51 & 3.86 \\ \hline
\end{tabular}
\begin{tabular}{||l|l|l||}
\multicolumn{3}{c}{$\bar{B}^0 \to \Xi_{cs} \bar{\Lambda}_c + X$}\\ \hline
   A &    B &    C \\ \hline
2.09 & 1.00 & 0.31 \\ \hline
\end{tabular}

\vspace{1cm}

\begin{tabular}{||l|l|l||}
\multicolumn{3}{c}{$B \to$ baryon + antibaryon}           \\ \hline
   A &    B &    C \\ \hline
8.89 & 5.51 & 4.17 \\ \hline
\end{tabular}

\end{center}

\newpage

\noindent {\bf Table 8}. $B$-meson branching ratios (${\cal{B}}r_{ij}
\equiv \Gamma_{ij} / \Gamma_B^{tot}$ given in $\%$) calculated for
exclusive and inclusive semileptonic charmed decays of the $B$-meson.

\begin{center}

\begin{tabular}{||l||lll||l||l||} \hline
Decay Mode & A& B& C& average & Exp. data \\ \hline
${\cal{B}}r(B \to X_c \ell \nu_{\ell})$ &10.35&11.24&11.59&$11.06 \pm
0.64$ &$10.90 \pm 0.46 ~ {\cite{NEU96}}$  \\ \hline
${\cal{B}}r(B \to X_c \tau \nu_{\tau})$ & 2.28& 2.34& 2.37&$ 2.33 \pm
0.05$ &$2.60 \pm 0.32 ~ {\cite{SKW95}}$   \\ \hline
${\cal{B}}r(B \to D e \nu_e)$           & 1.81& 1.93& 1.53&$ 1.76 \pm
0.21$ &$ 1.75 \pm 0.43 ~ {\cite{PDG96}}$  \\ \hline
${\cal{B}}r(B \to D^* e \nu_e)$         & 5.48& 5.58& 4.75&$ 5.27 \pm
0.45$ &$ 4.93 \pm 0.42 ~ {\cite{PDG96}}$  \\ \hline
\end{tabular}

\end{center}

\vspace{3cm}

\noindent {\bf Table 9}. The same as in Table 8, but for various
exclusive non-leptonic charmed decays of the $B$-meson.

\begin{center}

\begin{tabular}{||l||lll||l||l||} \hline
Decay Mode & A& B& C& average & Exp. data   \\ \hline
${\cal{B}}r(\bar{B}^0 \to D^+ \pi^-)$           & 0.40& 0.46& 0.34&$0.40
\pm 0.07$&$0.31 \pm 0.04 ~ {\cite{BHP96}}$  \\ \hline 
${\cal{B}}r(\bar{B}^0 \to D^+ \rho^-)$          & 0.89& 1.01& 0.74&$0.88
\pm 0.14$&$0.84 \pm 0.17 ~ {\cite{BHP96}}$  \\ \hline
${\cal{B}}r(\bar{B}^0 \to D^{*+} \pi^-)$        & 0.41& 0.46& 0.34&$0.40
\pm 0.05$&$0.28 \pm 0.04 ~ {\cite{BHP96}}$  \\ \hline 
${\cal{B}}r(\bar{B}^0 \to D^{*+} \rho^-)$       & 1.07& 1.17& 0.89&$1.04
\pm 0.14$&$0.73 \pm 0.15 ~ {\cite{BHP96}}$  \\ \hline
${\cal{B}}r(\bar{B}^0 \to D^+ D^-_s)$           & 1.53& 1.63& 1.02&$1.39
\pm 0.33$&$0.74 \pm 0.28 ~ {\cite{BHP96}}$  \\ \hline 
${\cal{B}}r(\bar{B}^0 \to D^+ D^{*-}_s)$        & 1.27& 1.33& 1.08&$1.23
\pm 0.13$&$1.14 \pm 0.50 ~ {\cite{BHP96}}$  \\ \hline
${\cal{B}}r(\bar{B}^0 \to D^{*+} D^-_s)$        & 1.08& 1.12& 0.92&$1.04
\pm 0.11$&$0.94 \pm 0.33 ~ {\cite{BHP96}}$  \\ \hline 
${\cal{B}}r(\bar{B}^0 \to D^{*+} D^{*-}_s)$     & 3.31& 3.42& 2.85&$3.19
\pm 0.30$&$2.00 \pm 0.64 ~ {\cite{BHP96}}$  \\ \hline
${\cal{B}}r(B \to charmed ~ baryons)$           & 8.79& 5.15& 3.77&$5.9 
\pm 2.6$&$6.4  \pm 1.1 ~ {\cite{PDG96}}$    \\
                                                &     &     &     &
        &$7.1  \pm 1.6 ~ {\cite{BHP96}}$    \\ \hline
$charm ~ counting$                              & 1.22& 1.20& 1.20&$1.20
\pm 0.01$&$1.16 \pm 0.05 ~ {\cite{CLEO_C}}$ \\
                                                &     &     &     &
        &$1.23 \pm 0.07 ~ {\cite{ALEPH_C}}$ \\ \hline
\end{tabular}

\end{center}

\vspace{3cm}

\noindent {\bf Table 10}. Inclusive $B$-meson branching ratios
(${\cal{B}}r_{ij} \equiv \Gamma_{ij} / \Gamma_B^{tot}$ given in $\%$) 
corresponding to the elementary transitions $b \to c \ell \nu_{\ell}$,
$b \to c \bar{c}s$ and $b \to c \bar{u}d$.

\begin{center}

\begin{tabular}{||l||lll||l||l||} \hline
Decay Mode & A& B& C& average & exp. data \\ \hline
${\cal{B}}r(b \to c \ell \nu_{\ell})$ &22.9&24.8&25.5&$24.4 \pm 1.3$
&$23.08 \pm 1.46 ~ \cite{L397}$ \\ \hline
${\cal{B}}r(b \to c \bar{c}s)$        &25.1&23.1&22.7&$23.6 \pm 1.3$
&$23.9 \pm 3.8 ~ \cite{BHP96}$  \\ \hline
${\cal{B}}r(b \to c \bar{u}d)$        &49.0&49.3&49.1&$49.1 \pm 0.2$
&                               \\ \hline
\end{tabular}

\end{center}

\newpage

\begin{figure}

\vspace{-5cm}

\epsfig{file=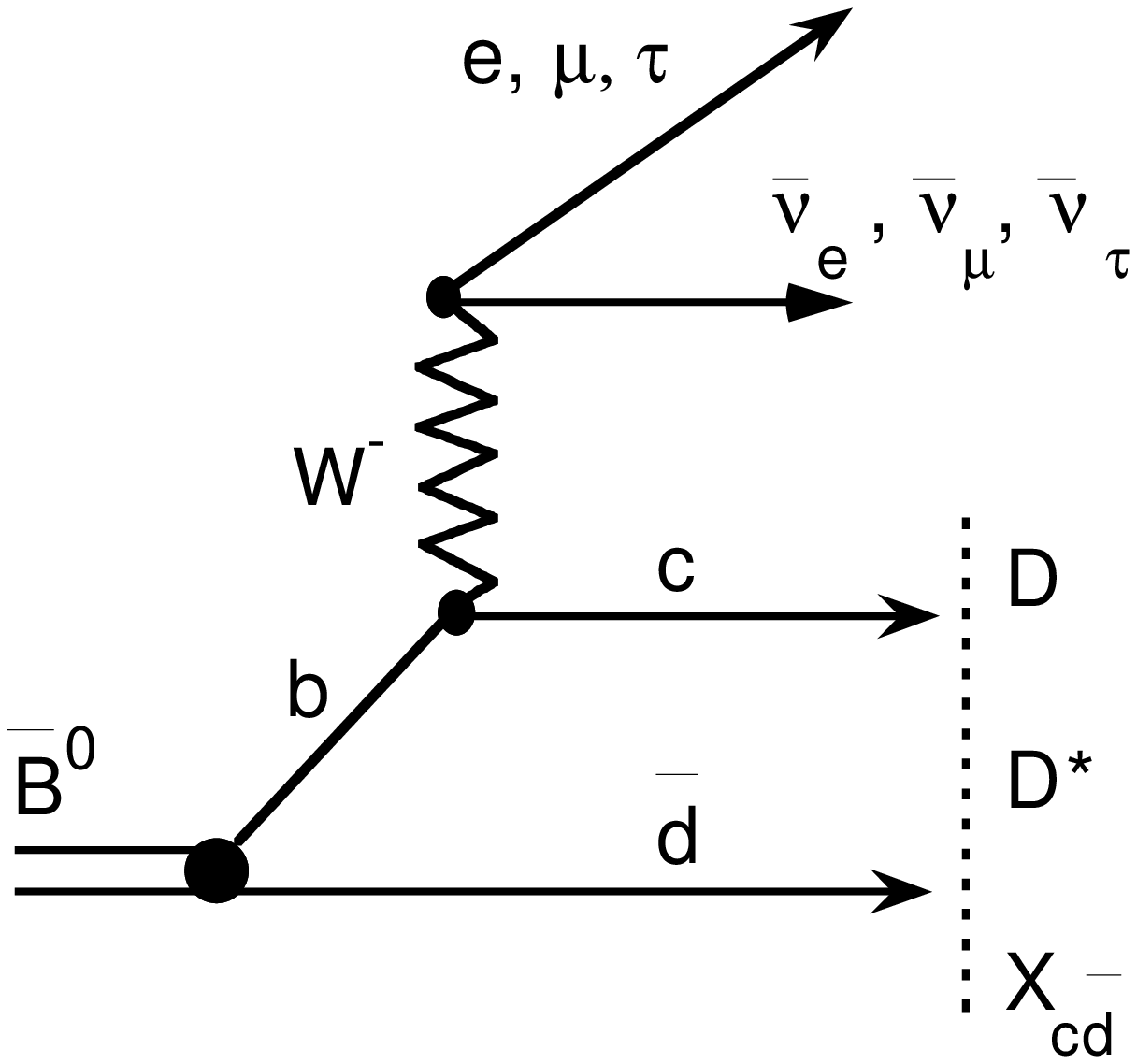}

\noindent {\bf Figure 1}. Semileptonic $\bar{B}^0$-meson decay modes. The
symbol $X_{q \bar{q}}$ stands for mesons with the flavour content of a $q
\bar{q}$ pair produced in the continuum.

\end{figure}

\newpage

\begin{figure}

\vspace{-2cm}

\epsfig{file=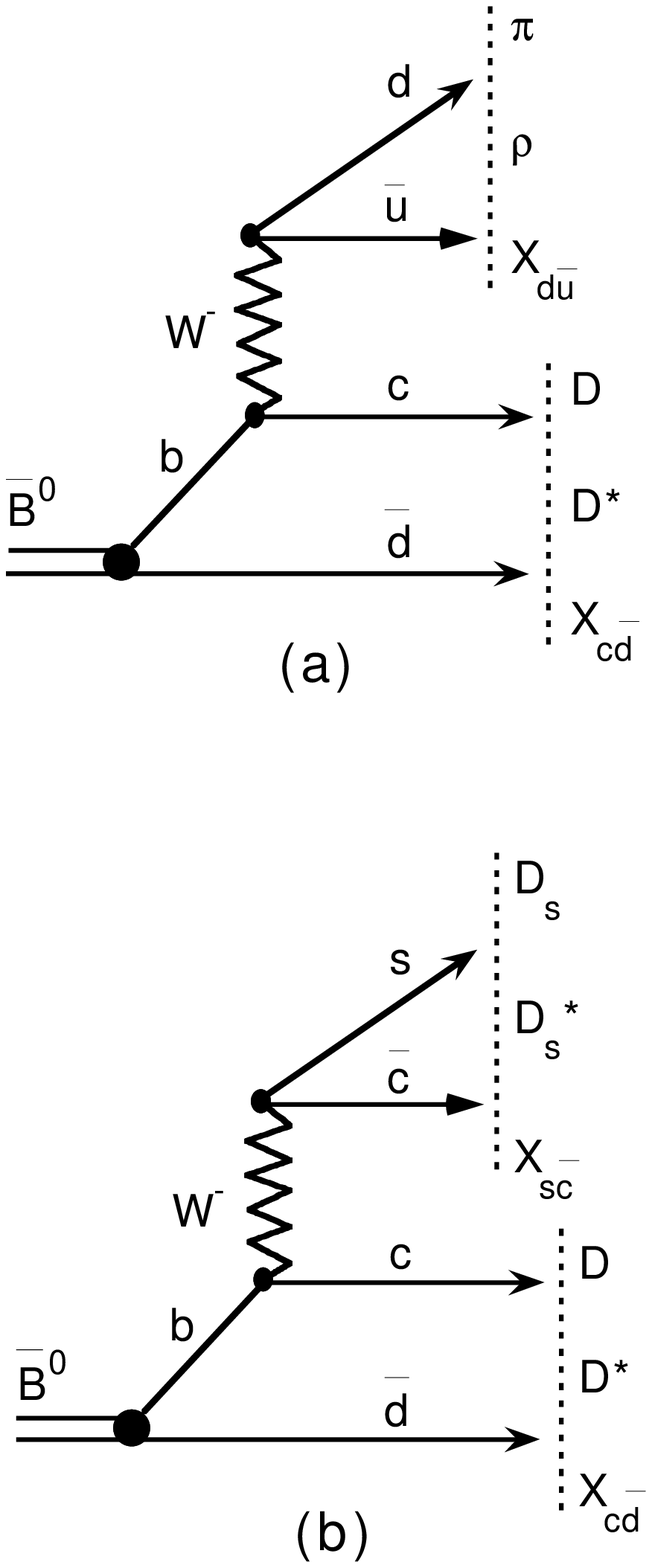}

\vspace{-1cm}

\noindent {\bf Figure 2}. Non-leptonic {\em external} $\bar{B}^0$-meson
decay modes, corresponding to $W^- \to \bar{u} d$ (a) and $W^- \to
\bar{c} s$ (b), respectively.

\end{figure}

\newpage

\begin{figure}

\vspace{-2cm}

\epsfig{file=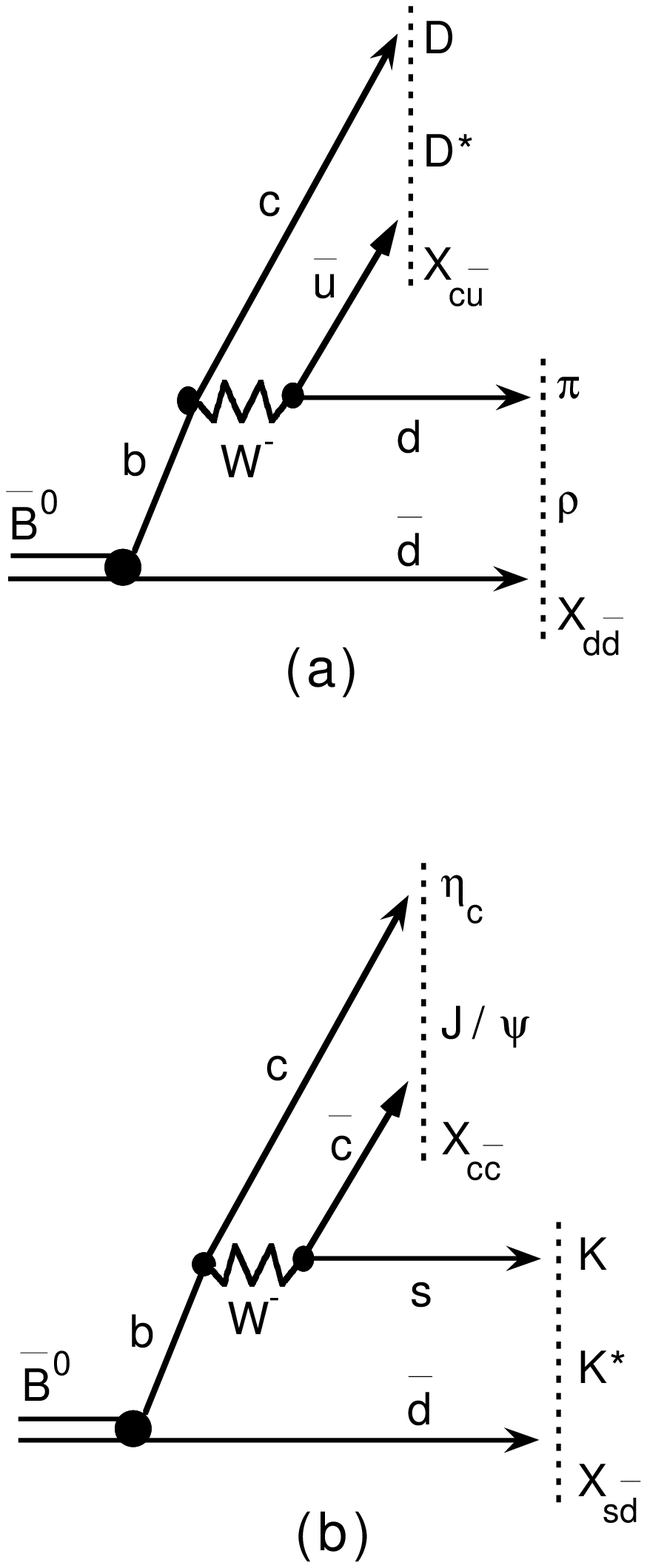}

\vspace{-1cm}

\noindent {\bf Figure 3}. Diagrams of {\em internal} non-leptonic
$\bar{B}^0$-meson decays into heavy mesons in case of the $W^- \to
\bar{u} d$ transition (a) and $W^- \to \bar{c} s$ transition (b).

\end{figure}

\newpage

\begin{figure}

\vspace{-5cm}

\epsfig{file=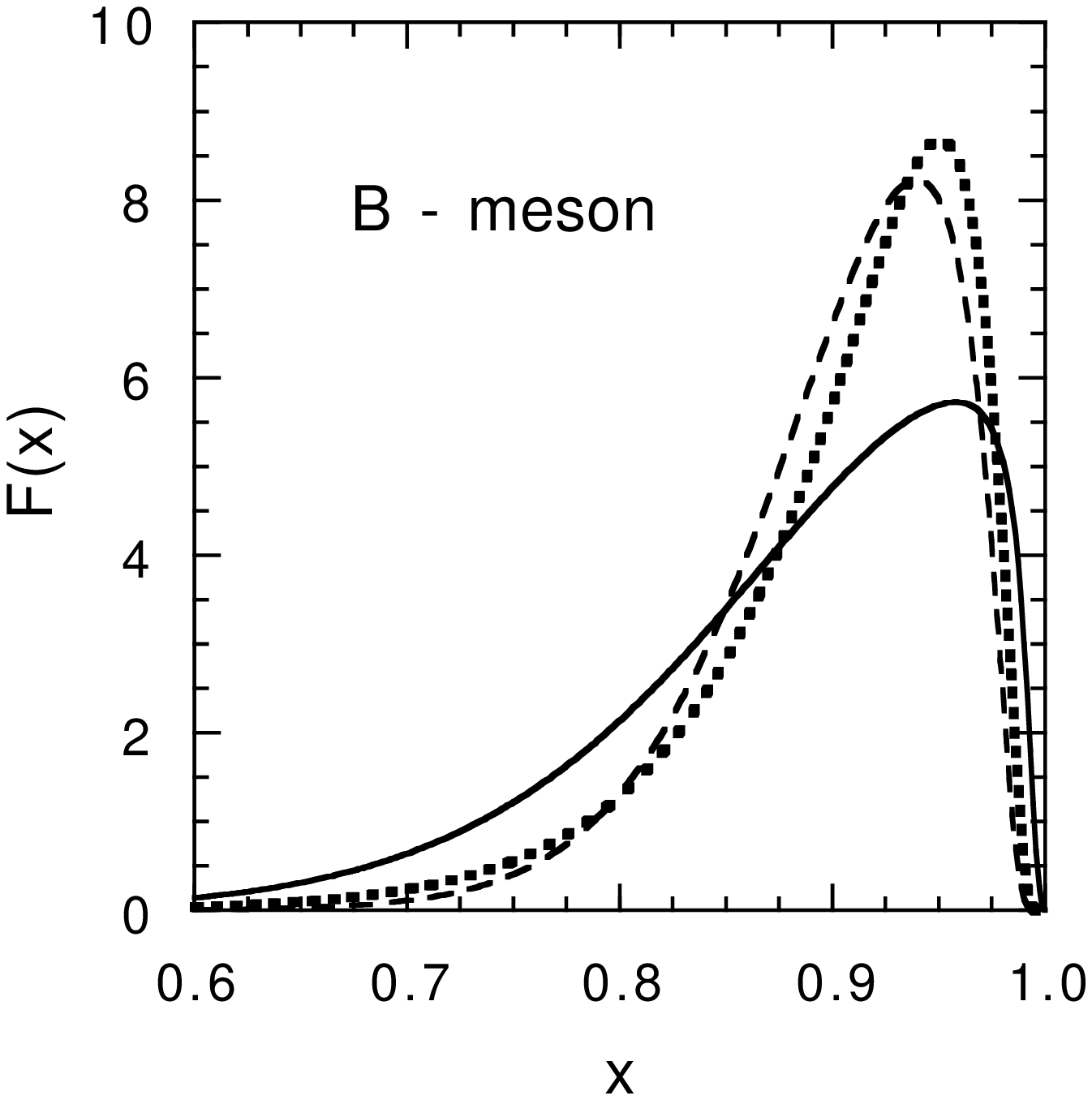}

\noindent {\bf Figure 4}. The distribution function $F(x)$ of a $b$-quark
inside the $B$-meson (Eq. (\ref{2.21})) versus the $LF$ momentum
fraction $x$. The dotted line is the result obtained using the
phenomenological ans\"atz of Eq. (\ref{2.11}). The solid and dashed
lines correspond to the calculation performed with $\chi_{GI}^{LF}$ and
$\chi_{NR}^{LF}$, obtained from the solution of Eq. (\ref{2.17}) with the
relativized interaction of Ref. \cite{GI85} and the solution of Eq.
(\ref{2.19}) with the non-relativistic potential of Ref. \cite{NCS92},
respectively.

\end{figure}

\newpage

\begin{figure}

\vspace{-5cm}

\epsfig{file=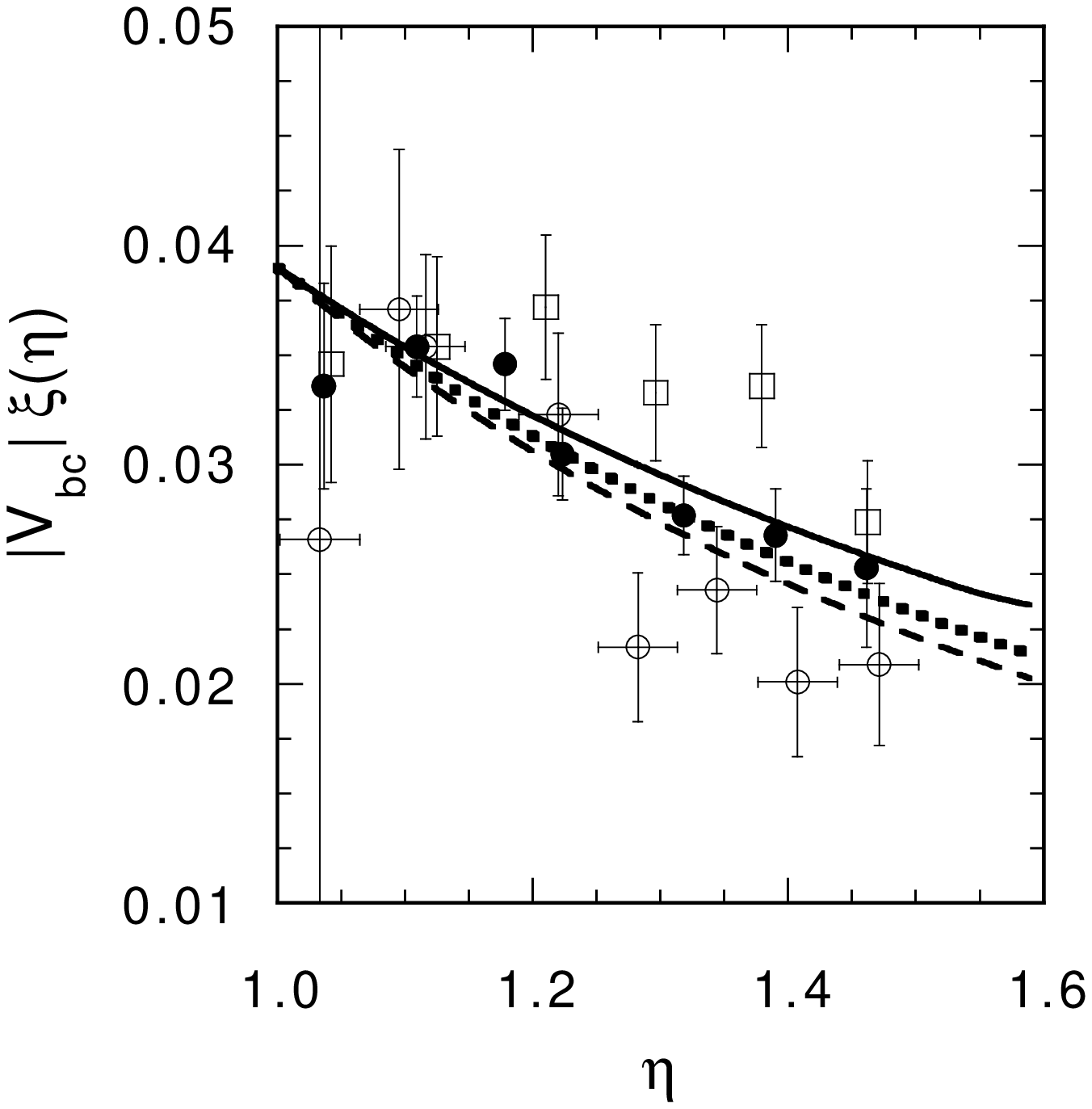}

\noindent {\bf Figure 5}. The $IW$ form factor $\xi(\eta)$, times
$|V_{bc}|$, as a function of the recoil $\eta$. The open dots, full dots
and squares correspond to the experimental data of Refs. \cite{ARGUS},
\cite{CLEOII}, \cite{ALEPH}, respectively. The dotted line is the result
of the calculations of Eq. (\ref{2.12}), times $|V_{bc}| = 0.0390$
\cite{NEU96}, obtained using the phenomenological ans\"atz of Eq.
(\ref{2.11}). The dashed and solid lines correspond to the results of
the calculations of Eq. (\ref{2.7}), times $|V_{bc}| = 0.0390$
\cite{NEU96}, obtained using the $LF$ wave functions $\chi_{NR}^{LF}$
and $\chi_{GI}^{LF}$, respectively.

\end{figure}

\newpage

\begin{figure}

\vspace{-2cm}

\epsfig{file=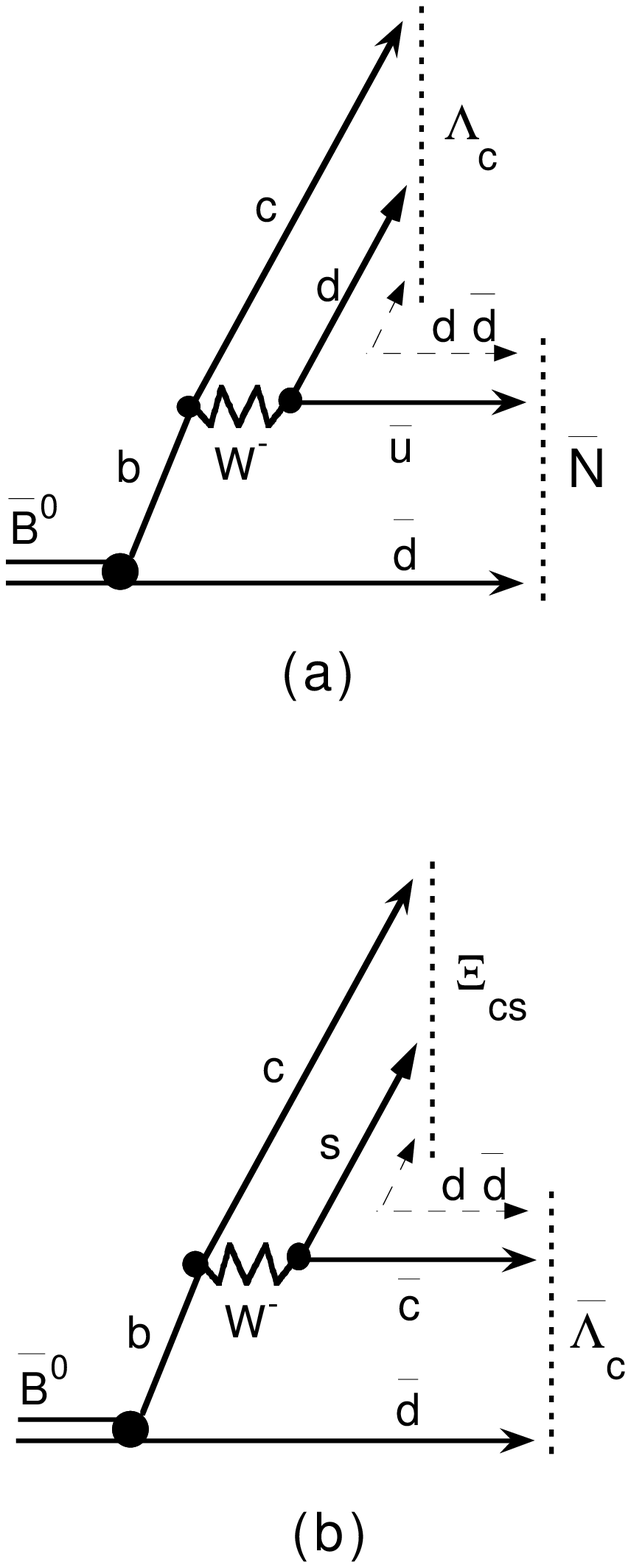}

\vspace{-1cm}

\noindent {\bf Figure 6}. (a) {\em Internal} non-leptonic decays of the
$\bar{B}^0$-meson into a diquark and an anti-diquark with the creation of
one more $q \bar{q}$ light-quark pair in case of the $W^- \to \bar{u} d$
transition. (b) The same as in (a), but for the $W^- \to \bar{c} s$
transition.

\end{figure}

\end{document}